\documentclass[pre,amsmath,amsfonts,nofootinbib,floatfix,twocolumn]{revtex4}

\newcommand{\PRE}[1]{#1}
\newcommand{\PREdel}[1]{}

\usepackage{subfigure}

\usepackage{graphicx}

\graphicspath{{Figures/}}

\newcommand{\figurewidth}{\columnwidth}
\newcommand{\subfigurewidth}{0.5\textwidth}

\usepackage[colorlinks,bookmarks]{hyperref}

\usepackage{dcolumn}
\newcolumntype{=}{D{=}{{}={}}{-1}}
\newcolumntype{,}{D{,}{{}={}}{-1}}
\newcolumntype{|}{D{|}{{}\quad{}}{-1}}

\usepackage[mediumspace,squaren,textstyle]{SIunits}
\addunit{\Equivalent}{equiv}

\begin{document}

\title{Phase field modeling of electrochemistry I: Equilibrium}

\author{J. E. Guyer}
\email{guyer@nist.gov}
\author{W. J. Boettinger}
\email{william.boettinger@nist.gov}
\author{J. A. Warren}
\email{jwarren@nist.gov}
\affiliation{Metallurgy Division, Materials Science and Engineering 
Laboratory, National Institute of Standards and Technology, 
Gaithersburg, MD  20899}

\author{G. B. McFadden}
\email{mcfadden@nist.gov}
\affiliation{Mathematical and Computational Sciences Division, 
Information Technology Laboratory, 
National Institute of Standards and Technology, 
Gaithersburg, MD  20899}

\begin{abstract}
A diffuse interface (phase field) model for an electrochemical system is
developed.  We describe the minimal set of components needed to model an
electrochemical interface and present a variational derivation of the governing
equations.  With a simple set of assumptions: mass and volume constraints,
Poisson's equation, ideal solution thermodynamics in the bulk, and a simple
description of the competing energies in the interface, the model captures the
charge separation associated with the equilibrium double layer at the
electrochemical interface.  The decay of the electrostatic potential in the
electrolyte agrees with the classical Gouy-Chapman and Debye-H\"uckel theories.
We calculate the surface \PRE{free} energy, surface charge, and differential capacitance as
functions of potential and find qualitative agreement between the model and
existing theories and experiments.  In particular, the differential capacitance
curves exhibit complex shapes with multiple extrema, as exhibited in many
electrochemical systems.
\end{abstract}


\date{\today} 

\maketitle

\section{Introduction}

We develop a phase field model of an electrochemical system.  The method employs
a phase field variable,  which is a function of position and
time, to describe whether the material is one phase or another (\emph{i.e.}, 
the electrode or electrolyte).  The behavior of
this variable is governed by a partial differential equation (PDE) that is
coupled to the relevant transport equations for the material.  The interface
between the phases is described by smooth but highly localized changes of this
variable.  This approach avoids the mathematically difficult problem of applying
boundary conditions at an interface whose location is part of the unknown
solution.  The phase field method is powerful because it easily treats complex
interface shapes and topology changes.  One long range goal of the approach is
to treat the complex geometry, including void formation, that occurs during
plating in vias and trenches for on-chip metallization \cite{Josell:2001}.

\PRE{Early models of the electrochemical interface, developed by Gouy, Chapman,
and Stern} \cite{Grahame:1947,Vetter:1967,BockrisReddy:2nd,Bard:2nd} \PRE{focused
on the distribution of charges in the electrolyte.  These models, which assume
that the charges have a Boltzmann distribution and are subject to Poisson's
equation, are summarized in Appendix~\ref{sec:Classical:GouyChapmanStern}.  More
recently, density functional models have been applied to the equilibrium
electrochemical interface} \cite{Goodisman:1987,Tang:1992}.  \PRE{These atomic
scale models describe the electrolyte with distribution functions which have
maxima at the positions of the atoms and take the electrode to be a hard,
idealized surface.  The equilibrium distribution of electrons and ions are
computed in these models and relationships between potential, charge, surface
free energy, and capacitance are obtained.  No kinetic modelling is performed in
these papers.  Phase field models can be viewed as a mean field approximation of
atomic scale density functional theories}
\cite{BoettingerReview:2002,McFaddenReview:2002,Harrowell:1987}, \PRE{and the
two methods often make similar predictions.}

The phase field method has been used widely for solidification
\cite{BoettingerReview:2002, McFaddenReview:2002}.  The present approach is
motivated by the mathematical analogy between the governing equations of
solidification dynamics and electroplating dynamics.  For example, the solid-melt
interface is analogous to the electrode-electrolyte interface.  The various
overpotentials of electrochemistry have analogies with the supercoolings of alloy
solidification: diffusional (constitutional), curvature and interface attachment.
Dendrites can form during solidification and during electroplating.  It is not
surprising, however, that we find significant differences between the two
systems.  The crucial presence of charged species in electrochemistry leads to
rich interactions between concentration, electrostatic potential, and phase
stability.

We first pick a minimal set of components required to describe the possible
composition variations from the electrode to the electrolyte through the
electrochemical interface.  Next we define concentration and mole fraction
variables.  Then a variational principle is used to establish a set of PDE's
that govern equilibrium interfaces.  In a second paper \cite{ElPhF:kinetics}, we
explore dynamic solutions to the phase field equations.

Given a minimal set of assumptions, the model predicts the charge separation
associated with the equilibrium double layer at the electrochemical interface.
Changes in surface potential induce changes in surface \PRE{free} energy
(electrocapillary curves), surface charge, and differential capacitance.  The
decay length of charge in the electrolyte as a function of electrolyte
concentration is consistent with the Debye-H\"uckel theory.  The parameters of
the phase field model are related to the physical parameters of traditional
electrochemistry.  The results are compared to the classic Gouy-Chapman-Stern
model of the electrified interface as well as to
experimental capacitance curves \cite{Valette:1973,Valette:1982}.

\section{Model Formulation}

\subsection{Choice of Components, Phases, and Molar Volumes}

To treat an electrochemical system, we consider a set of components consisting of
cations, anions, and electrons.  We will refer to the metallic conducting
electrode phase as \Electrode\ and the ionic conducting electrolyte phase as
\Electrolyte.  In this diffuse interface model, the concentrations of the
components will vary smoothly through the interfacial region.  The electrode will
be considered \PRE{a phase of primarily} two components: electrons
\Electron{} (component \#1) and cations \Cation{+\NumberCa} (component \#2).  The
electrolyte will be considered \PRE{a phase of primarily} three
components; more noble cations \Cation{+\NumberCa} (component \#2), less noble
cations \Otherion{+\NumberOt} (component \#3), and anions \Anion{-\NumberAn}
(component \#4).  \PRE{A model aqueous electrolyte can be considered as a special
case by setting \( \NumberOt = 0 \).} The primary charge transfer reaction for
this system is
\begin{equation}
    \Cation{+\NumberCa}\left(\Electrolyte\right)
    + \NumberCa \,\Electron\left(\Electrode\right)
    \rightleftharpoons \Cation{}\left(\Electrode\right).
    \label{eq:HalfCell}
\end{equation}

\PREdel{For the solution behavior,} All ions are treated as substitutional
species with identical partial molar volumes
\PartialMolarVolume{\Substitutional}.  Electrons are treated as an interstitial
species with zero partial molar volume.  The interstitial nature is necessary to
recover ohmic behavior in the electrode phase.  \PRE{In reality, the partial
molar volumes of the substitutional components should depend at least on the
species and phase, but this variation leads to deformation and
flow, which are not the focus of this work.} We employ mole fractions \Fraction{j}
of each component \( j \) that have the conventional definition, such that
\begin{equation}
    \sum_{j=1}^{\Components} \Fraction{j} = 1,
    \label{eq:MoleFraction}
\end{equation}
where \( \Components = 4 \) for the four-component system we consider in this
paper.  The molar volume varies because the mole fractions of the components vary
from one phase, through the interface, and into the other phase.  At each point
it is given by
\begin{equation}
    \MolarVolume = \sum_{j=1}^{\Components} \PartialMolarVolume{j} \Fraction{j}
     = \PartialMolarVolume{\Substitutional} \sum_{j=2}^{\Components} \Fraction{j},
\end{equation}
where \PartialMolarVolume{j}\ is the partial molar volume of each component \( j \), \(
\sum_{j=1}^{\Components} \) is the sum over all components, and \(
\sum_{j=2}^{\Components} \) is the sum over all substitutional components.  The 
mole per volume concentrations are defined as \( \Concentration{j} \equiv 
\Fraction{j} / \MolarVolume \), such that
\begin{equation}
    \sum_{j=1}^{\Components} \PartialMolarVolume{j}\Concentration{j}
    = \PartialMolarVolume{\Substitutional} \sum_{j=2}^{\Components} \Concentration{j}
    = 1.
    \label{eq:VolumeConstraint}
\end{equation}
Picking one component \( j = \Components \) with \(
\PartialMolarVolume{\Components} = \PartialMolarVolume{\Substitutional} \neq 0
\), we can always express its concentration in terms of the others as \(
\Concentration{\Components} = 1/\PartialMolarVolume{\Substitutional} -
\sum_{j=2}^{\Components-1} \Concentration{j} \).

\subsection{Equilibrium}

We propose a Helmholtz free energy for an isothermal system of charged
components
\begin{multline}
    \Helmholtz
    \left(
	\Phase,\Concentration{1}\ldots\Concentration{\Components},\Potential
    \right)
    \\
    = \int_{\Volume} 
    \left(
        \HelmholtzPerVol
        \left(
            \Phase,\Concentration{1}\ldots\Concentration{\Components}
        \right)
	+ \frac{\Gradient{\Phase}}{2}\abs{\nabla\Phase}^{2}
	+ \frac{1}{2} \ChargeDensity \Potential
    \right)
    d\Volume
    \label{eq:HelmholtzFromElectrostatics}
\end{multline}
integrated over the volume \Volume, where \HelmholtzPerVol\ is the Helmholtz
free energy per unit volume, \Phase\ is the phase field variable, \Potential\ is
the electrostatic potential, \Gradient{\Phase} is the gradient energy
coefficient for the phase field,
\begin{equation}
    \ChargeDensity \equiv \Faraday
    \sum_{j=1}^{\Components}\Valence{j}\Concentration{j}    
    \label{eq:ChargeDensity}
\end{equation}
is the charge density, \Valence{j} is the valence of component \( j \)
(\Equivalent\per\mole\ or charge units per ion), and \Faraday\ is
Faraday's constant.  The first term in the integral of
Eq.~\eqref{eq:HelmholtzFromElectrostatics} represents the energy
density of a system without gradients and with no charge
interactions; the second represents the gradient energy without
electrostatic effects; and the third represents the electrostatic
energy.

There are three constraints on the field variables of this system.  The total number of
moles \( \Number_j \) of each species \( j \) must be conserved over the volume
\Volume
\begin{equation}
    \Number_j 
    = \int_{\Volume} \Concentration{j} \, d\Volume 
    = \Volume \bar{\Concentration{j}},
    \qquad j = 1\ldots\Components
    \label{eq:MassConservation}
\end{equation}
where \( \bar{\Concentration{j}} \) is the average concentration of species \( j \).
In addition, Eq.~\eqref{eq:VolumeConstraint} and
Poisson's equation
\begin{equation}
    \nabla\cdot\left[\Dielectric(\Phase)\nabla\Potential\right] 
    + \ChargeDensity = 0
    \label{eq:Poisson}
\end{equation}
must be satisfied at every point in the system.  \( \Dielectric(\Phase) \) is the
electrical permittivity, whose value \PRE{we take to explicitly depend on the
phase.  Since the phase field is coupled to the other variables, there is also an
implicit dependence on the concentrations of different species in the
electrolyte.}

We note that by invoking Poisson's equation \eqref{eq:Poisson}, the electrostatic
energy term in Eq.~\eqref{eq:HelmholtzFromElectrostatics}, given by \(
\frac{1}{2} \ChargeDensity \Potential \), could equivalently be represented as \(
\frac{1}{2}\Displacement\cdot\ElectricField \), one half the scalar product of
the displacement and the electric field of electromagnetic theory, or as \(
\frac{\Dielectric}{2}\abs{\nabla\Potential}^{2} \).  In this third form, we see
that the electrostatic energy density is completely analogous to the phase field
gradient energy density.

We perform a variational analysis on the resulting Lagrangian
\begin{align}
    \Lagrangian
    = \Helmholtz 
    & {} - \int_{\Volume} \Lagrange{\Volume}(\vec{\Position})
    \left(
        \sum_{j=1}^{\Components} \PartialMolarVolume{j} \Concentration{j}
	-1
    \right)
    \, d\Volume
    \nonumber \\
    & {} - \sum_{j=1}^{\Components} \Lagrange{j} \int_{\Volume} 
	\left(\Concentration{j} - \bar{\Concentration{j}}\right) 
	\, d\Volume
    \nonumber \\
    & {} - \int_{\Volume} \Lagrange{\Potential}(\vec{\Position})
	\left\{
	    \nabla\cdot\left[
		\Dielectric(\Phase)\nabla\Potential
		\right] + \ChargeDensity
	\right\} d\Volume
    \label{eq:Lagrangian}
\end{align}
where we have introduced the Lagrange multipliers \(
\Lagrange{\Volume}(\vec{\Position}) \), \Lagrange{j}, and \(
\Lagrange{\Potential}(\vec{\Position}) \) for the constraints
\eqref{eq:VolumeConstraint}, \eqref{eq:MassConservation}, and \eqref{eq:Poisson}
(note that the requirement that Eqs.~\eqref{eq:VolumeConstraint} and
\eqref{eq:Poisson} be satisfied everywhere in the system means that
\Lagrange{\Volume} and
\Lagrange{\Potential} must be fields).  At equilibrium, each of
the variations of \Lagrangian\ must be independently zero,
\begin{subequations}
    \begin{align}
	\frac{\Variation \Lagrangian}{\Variation \Concentration{j}} 
	&= 0
	= \frac{\partial \HelmholtzPerVol}{\partial \Concentration{j}}
	+ \frac{1}{2}\Faraday\Valence{j}\Potential
	- \Lagrange{j}
	- \Faraday\Valence{j}\Lagrange{\Potential}
	- \Lagrange{\Volume}\PartialMolarVolume{j}
	\label{eq:Lagrange:Concentration} 
	\\
	& \qquad\qquad\qquad\qquad\qquad\qquad 
	j = 1\ldots\Components
	\nonumber \\
	\frac{\Variation \Lagrangian}{\Variation \Phase} 
	&= 0
	= \frac{\partial \HelmholtzPerVol}{\partial \Phase}
	- \Gradient{\Phase} \nabla^{2} \Phase
	+ \Dielectric'(\Phase) \nabla\Lagrange{\Potential} \nabla\Potential
	\label{eq:Lagrange:Phase} \\
	\frac{\Variation \Lagrangian}{\Variation \Potential} 
	&= 0
	= \frac{1}{2}\ChargeDensity
	- \nabla\cdot\left[
	    \Dielectric(\Phase)\nabla\Lagrange{\Potential}
	\right].
	\label{eq:Lagrange:Potential} 
    \end{align}%
    \label{eq:Lagrange}%
\end{subequations}%
Using Poisson's equation to eliminate \ChargeDensity\ in
Eq.~\eqref{eq:Lagrange:Potential}, we determine that \( \Lagrange{\Potential} =
-\Potential / 2 \).  Here we have assumed that the natural boundary conditions \(
0 = \partial \Phase / \partial \vec{\Normal} = \partial \Potential / \partial
\vec{\Normal} = \partial \Lagrange{\Potential} / \partial \vec{\Normal} \) hold
on the boundary of \Volume.  Substituting this result into
Eq.~\eqref{eq:Lagrange:Concentration},
\begin{equation}
    \Lagrange{j} 
    = \frac{\partial \HelmholtzPerVol}{\partial \Concentration{j}}
    + \Faraday\Valence{j}\Potential
    - \Lagrange{\Volume}(\vec{\Position})\PartialMolarVolume{j}.
    \qquad j = 1\ldots\Components
    \label{eq:Lagrange:Concentration:Substituted}
\end{equation}
Since \( \PartialMolarVolume{\Electron} = 0 \) and \( \PartialMolarVolume{j} =
\PartialMolarVolume{\Substitutional} \) for all substitutional species, we can
eliminate \( \Lagrange{\Volume}(\vec{\Position}) \).

We thus can summarize the equilibrium governing equations as
\begin{subequations}
    \begin{align}
	\Lagrange{j\Solvent{}}
	&= \Lagrange{j}
	- \frac{\PartialMolarVolume{j}}{\PartialMolarVolume{\Substitutional}} 
	\Lagrange{\Solvent{}}
	\nonumber \\
	&= \left[
	    \frac{\partial \HelmholtzPerVol}{\partial \Concentration{j}}
	    + \Faraday\Valence{j}\Potential
	\right]
	- \frac{\PartialMolarVolume{j}}{\PartialMolarVolume{\Substitutional}}
	\left[
	    \frac{\partial \HelmholtzPerVol}{\partial \Concentration{\Solvent{}}}
	    + \Faraday\Valence{\Solvent{}}\Potential
	\right]
	\nonumber \\
	&= \text{constant}
	\qquad j = 1\ldots\Components-1
	\label{eq:Governing:Electrochemical} \\
	0 &=
	\frac{\partial \HelmholtzPerVol}{\partial \Phase}
	- \Gradient{\Phase} \nabla^{2} \Phase
	- \frac{\Dielectric'(\Phase)}{2} \left(\nabla\Potential\right)^{2}
	\label{eq:Governing:Phase} \\
	0 &= \nabla\cdot\left[
	    \Dielectric(\Phase)\nabla\Potential
	\right] + \ChargeDensity.
	\label{eq:Governing:Poisson}
    \end{align}
    \label{eq:Governing}
\end{subequations}

It is convenient to identify the \PRE{classical} chemical potentials
\begin{equation}
    \Chemical{j}{} = \frac{\partial \HelmholtzPerVol}{\partial 
\Concentration{j}}
    \qquad j = 1\ldots\Components
    \label{eq:Cchemical:Classic}
\end{equation}
and the \PRE{classical} electrochemical potentials
\begin{equation}
    \Electrochemical{j} 
    = \frac{\partial \HelmholtzPerVol}{\partial \Concentration{j}}
    + \Faraday\Valence{j}\Potential,
    \qquad j = 1\ldots\Components
    \label{eq:Electrochemical:Classic}
\end{equation}
such that Eq.~\eqref{eq:Governing:Electrochemical} becomes
\begin{equation}
    \Electrochemical{j}
    - \frac{\PartialMolarVolume{j}}{\PartialMolarVolume{\Substitutional}} 
    \Electrochemical{\Solvent{}} 
    = \text{constant}.
    \label{eq:Governing:Electrochemical:Classic}
\end{equation}
In Appendix~\ref{sec:FirstIntegral} we show that, in one dimension, 
\begin{equation}
   \Electrochemical{j} 
   = \Electrochemical{j}^\infty 
   + \PartialMolarVolume{j}
   \left(
       \frac{\Gradient{\Phase}}{2} \Phase_\Position^2 
      - \frac{\Dielectric(\Phase)}{2} \Potential_\Position^2
   \right),
   \qquad j = 1\ldots\Components
   \label{eq:Electrochemical:Interface}
\end{equation}
where \( \Electrochemical{j}^\infty \) is the electrochemical potential far from
the interface.  First, we note that far from the interface, where \( \nabla\Phase
= \nabla\Potential = 0 \), the value of the electrochemical potentials for each
component \( j \) are identical in each phase, in agreement with Gibbsean
thermodynamics \cite{Gibbs:06}
\begin{equation}
    \Electrochemical{j}^{\Electrode} = \Electrochemical{j}^{\Electrolyte}.
    \qquad j = 1\ldots\Components
    \label{eq:ElectrochemicalEquilibrium}
\end{equation}
Second, the electrochemical potential of \Electron\ is uniform throughout the
system.  For the substitutional components, the electrochemical potential varies
through the interface, even though the values in the bulk phases are equal.
Because \( \nabla\Potential = 0 \) far from the interface, in the absence of
external charges, Eq.~\eqref{eq:Governing:Poisson} requires that charge is zero
in the bulk phases, such that
\begin{equation}
    \sum_{j=1}^{\Components} \Valence{j} \Fraction{j}^{\Electrode} =
    \sum_{j=1}^{\Components} \Valence{j} \Fraction{j}^{\Electrolyte} = 0.
    \label{eq:ZeroCharge}
\end{equation}

\section{Interfacial Properties}
\label{sec:Interface}

From electrocapillary theory \cite{Grahame:1947}, we know that surface \PRE{free} energy
\SurfaceEnergy, excess charge on the electrode \( \SurfaceCharge^\Electrode \),
and differential capacitance \DifferentialCapacitance\ are related by
\begin{align}
    \SurfaceCharge^\Electrode 
    &= -\left(
	\frac{\partial\SurfaceEnergy}{\partial\Galvani{}}
    \right)_{\Chemical{j}{}}
    \label{eq:Adsorption:Classical} \\
    \DifferentialCapacitance
    &\equiv \left(
	\frac{\partial\SurfaceCharge^\Electrode}{\partial\Galvani{}}
    \right)_{\Chemical{j}{}}
    = -\left(
	\frac{\partial^2\SurfaceEnergy}{\partial{\Galvani{}}^2}
    \right)_{\Chemical{j}{}}
    \label{eq:DifferentialCapacitance}    
\end{align}
where \( \Galvani{} = \Potential^{\Electrode} - \Potential^{\Electrolyte} \) is
the applied potential difference across the interface.\footnote{In classical
electrochemistry, the actual difference in electrostatic potential across the
interface is called the ``inner'' or ``Galvani'' potential difference.} The
potential in the electrolyte far from the interface is \(
\Potential^{\Electrolyte} \) and that in the electrode is \(
\Potential^{\Electrode} \).  In a perfect conductor \( \Potential^{\Electrode} \)
is uniform throughout the phase.  The variation at constant chemical potential
(constant concentration) is difficult to perform with physical
electrode/electrolyte systems.  In general, this variation can only be performed
experimentally for an inert system such as a mercury electrode against an aqueous
electrolyte.  We seek to determine whether these relationships hold for our phase
field model. 

In Appendix~\ref{sec:SurfaceEnergy}, we derive the expression for the surface 
\PRE{free} energy
\begin{equation}
    \SurfaceEnergy 
    = \int_{-\infty}^{\infty}
    \left[ 
	\Gradient{\Phase} \Phase_\Position^2 
	- \Dielectric(\Phase) \Potential_\Position^2 
    \right] \, d\Position.
    \label{eq:SurfaceEnergy}
\end{equation}
The first term in the integrand represents the contributions of the
non-electrified interface.  The second term is the contribution of
electrostatics, such that the presence of electric fields always reduces the
surface \PRE{free} energy from its charge-free value.  In
Appendix~\ref{sec:SurfaceCharge} we obtain the relationship
\begin{equation}
    \SurfaceCharge^{\Electrode}
    = -\left(
	\frac{\partial \SurfaceEnergy}{\partial \Galvani{\Standard}}
    \right)_{\Chemical{j}{}}
    \label{eq:Adsorption}
\end{equation}
if \( \SurfaceCharge^{\Electrode} \) is defined by
\begin{equation}
    \SurfaceCharge^\Electrode
    \equiv \int_{-\infty}^\infty \Interpolate(\Phase) \ChargeDensity \, d\Position,
    \label{eq:SurfaceCharge:Electrode}
\end{equation}
which is a completely reasonable definition of the surface charge on the
electrode.  \( \Interpolate(\Phase) \) is an interpolation function that will be
described in Section~\ref{sec:Parameters:Function}; in short, \(
\Interpolate(\Phase) = 1 \) in the electrode and \( \Interpolate(\Phase) = 0 \)
in the electrolyte.  The notation \Galvani{\Standard} differs from that in
Eqs.~\eqref{eq:Adsorption:Classical} and \eqref{eq:DifferentialCapacitance}.
\Galvani{\Standard} refers to a materials property of the electrode-electrolyte
system, as we will discuss in Section~\ref{sec:Parameters:Thermodynamics}.
Because we consider a non-inert electrode, we cannot vary the applied potential
without affecting the concentration in the electrolyte.  The variation considered
in Eq.~\eqref{eq:Adsorption} is actually a variation with respect to a changing
material property, rather than an applied potential.  The differential
capacitance is then defined to be
\begin{equation}
    \DifferentialCapacitance
    \equiv \left(
	\frac{\partial\SurfaceCharge^\Electrode}{\partial\Galvani{\Standard}}
    \right)_{\Chemical{j}{}}.
    \label{eq:DifferentialCapacitance:Ours}
\end{equation}
We note that a curved interface exhibits an additional relationship between
surface \PRE{free} energy, surface concentration, and the interfacial potential drop,
consistent with the Gibbs-Thomson effect \cite{GBMcF:GibbsThomson}.

\section{Thermodynamic Functions and Material Parameters}
\label{sec:Parameters}

To make the results of the phase field model more concrete and to permit 
numerical calculations, we must choose a particular form of the thermodynamic 
function and the materials parameters, at which point the model will 
be fully specified.

\subsection{Choice of Form of the Thermodynamic Function}
\label{sec:Parameters:Function}

For simplicity, we assume that the chemical part of the Helmholtz free energy per
unit volume is described by an \PRE{interpolation of two ideal solutions of the
components for the electrode and electrolyte}.
\begin{align}
    \HelmholtzPerVol
    \left(
	\Phase,\Concentration{1}\ldots\Concentration{\Components}
    \right)
    &= \frac{1}{\MolarVolume} \HelmholtzPerMol
    \left(
	\Phase,\Fraction{1}\ldots\Fraction{\Components}
    \right)
    \nonumber \\
    &= \sum_{j=1}^{\Components} \Concentration{j} 
	\left\{
	    \Chemical{j}{\Standard\Electrode}\Interpolate\left(\Phase\right)
	    + \Chemical{j}{\Standard\Electrolyte}
		\left[1-\Interpolate\left(\Phase\right)\right]
	\right.
	\nonumber \\
    &\qquad \left.
	\vphantom{\Chemical{j}{\Standard\Electrolyte}}
	    + \Gas\Temperature\ln\Concentration{j}\MolarVolume
	    + \Barrier{j} \DoubleWell\left(\Phase\right)
	\right\}
    \label{eq:HelmholtzPerVolume}
\end{align}
where \Chemical{j}{\Standard\Electrode} and \Chemical{j}{\Standard\Electrolyte}
are the chemical potentials of pure component \( j \) in the electrode (metal)
phase and the electrolyte phase respectively, \Gas\ is the molar gas constant,
and \Temperature\ is the temperature.  Following many phase field models of
solidification \cite{Wang:1993}, we use an interpolating function \(
\Interpolate\left(\Phase\right) = \Phase^{3}\left(6\Phase^{2} - 15\Phase +
10\right) \) to bridge between the descriptions of the two bulk phases and a
double-well function \( \DoubleWell\left(\Phase\right) =
\Phase^{2}\left(1-\Phase\right)^{2} \) with a barrier height of \Barrier{j} for
each component \( j \) to establish the metal/electrolyte interface.  The barrier
heights \Barrier{j} penalize interfaces which are too broad and the gradient
energy coefficient \Gradient{\Phase} penalizes interfaces which are too narrow.
The polynomials are chosen to have the properties that \( \Interpolate(0) = 0 \),
\( \Interpolate(1) = 1 \), \( \Interpolate'(0) = \Interpolate'(1) = 0 \), and \(
\DoubleWell'(0) = \DoubleWell'(1) = 0 \).  Other functions could be used.

For use in Eq.~\eqref{eq:Governing:Phase}, this free energy leads to
\begin{equation}
    \frac{\partial \HelmholtzPerVol}{\partial \Phase}
    = \Interpolate'\left(\Phase\right) 
    \sum_{j=1}^{\Components} \Concentration{j} \Delta\Chemical{j}{\Standard}
    + \DoubleWell'\left(\Phase\right)
    \sum_{j=1}^{\Components} \Concentration{j} \Barrier{j},
    \label{eq:Partial:Phase}
\end{equation}
where \( \Delta\Chemical{j}{\Standard} \equiv
\Chemical{j}{\Standard\Electrode}-\Chemical{j}{\Standard\Electrolyte} \).

The quantity \( \partial\HelmholtzPerVol/\partial\Concentration{j} \) is also
computed to give the electrochemical potentials,
\begin{multline}
	\Electrochemical{j} 		
	= \Chemical{j}{\Standard\Electrolyte}
	+ \Delta\Chemical{j}{\Standard}\Interpolate\left(\Phase\right)
	+ \Gas\Temperature
	    \ln\Concentration{j}\MolarVolume
	+ \Valence{j}\Faraday\Potential
	+ \Barrier{j} \DoubleWell\left(\Phase\right).
	\\
	\qquad j = 1\ldots\Components
	\label{eq:Electrochemical:Ideal}
\end{multline}
We note that the \Electrochemical{j} depend on all the \Concentration{j} 
through the molar volume \MolarVolume.

\subsection{Standard Chemical Potentials}
\label{sec:Parameters:Thermodynamics}

We require values for  the
parameters \( \Delta\Chemical{j}{\Standard} \) for each of the \Components\ 
species in our model in order to perform numerical calculations with the ideal 
solution thermodynamic model employed here.  Given these numbers, we
will know how the bulk electrode and electrolyte concentrations vary with the
potential difference across the interface.  From the equality of the bulk
electrochemical potentials Eq.~\eqref{eq:ElectrochemicalEquilibrium} and the
ideal solution form of the electrochemical potential
Eq.~\eqref{eq:Electrochemical:Ideal},
\begin{equation}
    \Galvani{}
    = -\frac{\Delta\Chemical{j}{\Standard}}{\Valence{j}\Faraday}
    + \frac{\Gas\Temperature}{\Valence{j}\Faraday}\ln
	\frac{\Fraction{j}^{\Electrolyte}}
	    {\Fraction{j}^{\Electrode}},
    \qquad j = 1\ldots\Components
    \label{eq:OurNernst}
\end{equation}
where the mole fractions are constrained by charge neutrality of the bulk phases
Eq.~\eqref{eq:ZeroCharge} and the ordinary definition of the mole fractions
Eq.~\eqref{eq:MoleFraction}.

One procedure to obtain information about the \( \Delta\Chemical{j}{\Standard}
\) is to specify one set of concentrations that are at equilibrium, \(
\Fraction{j}^{\Electrode\Standard} \) and \(
\Fraction{j}^{\Electrolyte\Standard} \).  With such a set we can only determine
three linear combinations of \( \Delta\Chemical{j}{\Standard} \).  Given these
three linear combinations, we cannot determine the potential difference across
the interface, but we can calculate how the bulk electrode and electrolyte
concentrations vary with changes in the potential difference across the
interface.  In other words, the potential difference can only be described with
respect to a reference electrode.  Although knowledge of the potential
difference between the phases is not necessary to describe the bulk equilibrium,
it is necessary if one is interested in modeling the charge distribution between
the electrode and electrolyte.  We will designate \Galvani{\Standard} as the
potential difference across the interface for the mole fractions \(
\Fraction{j}^{\Electrode\Standard} \) and \(
\Fraction{j}^{\Electrolyte\Standard} \), such that
\begin{equation}
    \Delta\Chemical{j}{\Standard} 
    = \Gas\Temperature\ln
    \frac{\Fraction{j}^{\Electrolyte\Standard}}
	{\Fraction{j}^{\Electrode\Standard}}
    - \Valence{j}\Faraday\Galvani{\Standard}.
    \qquad j = 1\ldots\Components
    \label{eq:DeltaStandard:Standard}
\end{equation}
Thus all four values of \( \Delta\Chemical{j}{\Standard} \) can be computed. 
The quantity \Galvani{\Standard} is a material property that is fixed for a 
given choice of electrode and electrolyte system.

In traditional electrochemistry, one takes the reference mole fractions of the
electroinactive species to be zero in one phase or the other, \emph{e.g.}, \(
\Fraction{\Electron}^{\Standard\Electrolyte} =
\Fraction{\Anion{-}}^{\Standard\Electrode} = 0 \).  One would then only equate
the electrochemical potentials between the bulk phases of the electroactive
species \Cation{+\NumberCa} and \Otherion{+\NumberOt}. In this case it can be shown 
that
\begin{equation}
    \EMF{\Cation{+\NumberCa}} - \EMF{\Otherion{+\NumberOt}}
    = - \left(
	\frac{\Delta\Chemical{\Cation{+\NumberCa}}{\Standard}}{\NumberCa\Faraday} 
	-\frac{\Delta\Chemical{\Otherion{+\NumberOt}}{\Standard}}{\NumberOt\Faraday} 
    \right),
    \label{eq:DeltaDelta}
\end{equation}
where the standard potentials \EMF{\Cation{+\NumberCa}} are
\EMF{\Otherion{+\NumberOt}} are obtained from a table of electromotive series. 
Eq.~\eqref{eq:DeltaDelta} would be adequate to describe the concentration
variations between the bulk phases.

For numerical purposes, we must assume small, but non-zero values for the
reference concentrations of the electroinactive species, equivalent to assuming
large (positive or negative), but finite, values for their standard potentials.
In reality these concentrations are not zero, as might be indicated by an
electrolyte with electronic
conductivity or an electrode with some anion solubility
\cite{Grahame:1947,Hart:1962,Egan:1985}.

In the remainder of the paper, we perform a detailed equilibrium analysis for an
electrolyte where the component \Otherion{} has no charge, like an aqueous system
(\Otherion{} = \Water) where dissociation of \Water\ is neglected.\footnote{The
dissociation of \Water\ could be handled by the addition of another component}
The lower density of charged species in an unsupported ``aqueous'' electrolyte
allows us to resolve the equilibrium interface more accurately than in a system
where all species can carry charge.  Our paper on the dynamic behavior of the
electrochemical phase field model \cite{ElPhF:kinetics} will treat the case of
\Otherion{} with charge.  The bulk standard state mole fractions are chosen for
an electrode of \Cation{} metal and an electrolyte of a solution containing
\unit{1}{\mole\per\liter} of \Cation{+2} and \Anion{-2} in a \Otherion{} solvent.
We take the partial molar volume of the ``substitutional'' components
(\Cation{+2}, \Anion{-2}, and \Otherion{}) as that for pure water; \emph{i.e.},
\( \PartialMolarVolume{\Substitutional} = \unit{0.018}{\liter\per\mole} \) or \(
1/\PartialMolarVolume{\Substitutional} = \unit{55.6}{\mole\per\liter} \).  The
voltage-independent portion of the chemical potential differences are give in
Table~\ref{tab:Parameters:Thermodynamic:Aqueous}.
\begin{table}[tbp]
	\centering
	\caption{Numerical values of the potential-independent portion of the
	chemical potential differences \( \Delta\Chemical{j}{\Standard}
	\).}
    \begin{ruledtabular}
	\begin{tabular}{cd}
	    & \multicolumn{1}{c}
		{$\ln (\Fraction{j}^{\Electrolyte\Standard} /
		    \Fraction{j}^{\Electrode\Standard})$}
	\\

	\hline

	\Electron
	& -13.41
	\\
	
	\Cation{+2}
	& -2.919
	\\
	
	\Anion{-2}
	& 9.798
	\\
	
	\Otherion{}
	& 13.78
	
    \end{tabular}
    \end{ruledtabular}    
	\protect\label{tab:Parameters:Thermodynamic:Aqueous}
\end{table}
The values for the electroinactive species are chosen to limit the corresponding
standard state mole fractions to \( \Fraction{\Electron}^{\Electrolyte\Standard}
= \Fraction{\Anion{-2}}^{\Electrode\Standard} =
\Fraction{\Otherion{}}^{\Electrode\Standard} = \power{10}{-6} \).

\begin{figure}[tbp]
    \centering
    \includegraphics[width=\figurewidth]{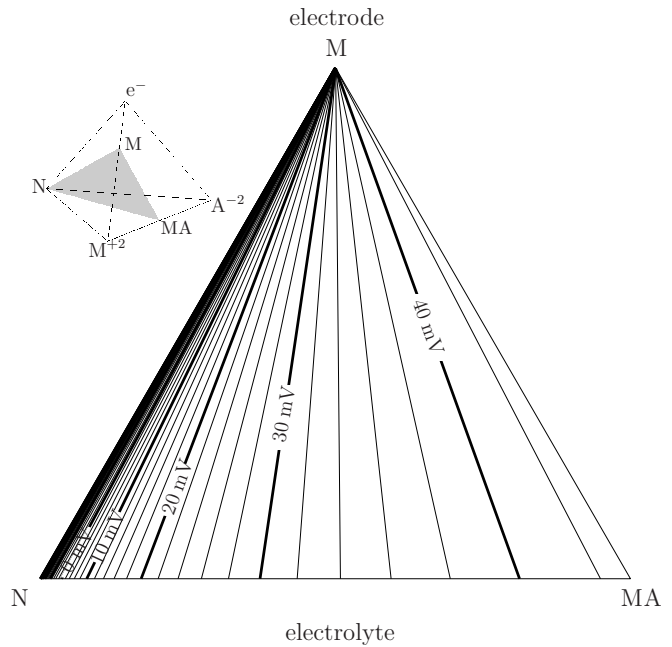}
    \caption{\PRE{Potential-composition} phase diagram for the parameters in
    Table~\ref{tab:Parameters:Thermodynamic:Aqueous}, \PRE{illustrating the bulk
    equilibrium between a \Cation{} electrode and a \Otherion{} electrolyte
    with dissolved \Cation{}\Anion{} salt}.  Tie-lines denote different values of the
    quantity \( (\Galvani{} - \Galvani{\Standard}) \).  The inset shows the
    position of this charge neutral phase diagram within the quaternary domain of
    the charged species.}
    \protect\label{fig:Nernst}
\end{figure}
Bulk charge neutrality can be invoked to transform the four mole fraction
variables \Fraction{\Electron}, \Fraction{\Cation{+2}}, \Fraction{\Anion{-2}},
and \Fraction{\Otherion{}} into mole fractions of three charge-neutral compounds,
\( \Fraction{\Cation{}} \equiv \frac{3}{2}\Fraction{\Electron} \), \(
\Fraction{\Cation{}\Anion{}} \equiv 2\Fraction{\Anion{-2}} \), and
\Fraction{\Otherion{}}.  We plot the equilibrium phase diagram in terms of these 
transformed mole fraction coordinates in Figure~\ref{fig:Nernst}. 
Equilibrium states exist only between \( \unit{-0.0898}{\volt} < \Galvani{} <
\unit{+0.0427}{\volt} \).  It can be seen that the electrode remains essentially
pure \Cation{} (\( \Fraction{\Cation{+2}}^{\Electrode}=1/3 \), \(
\Fraction{\Electron}^{\Electrode}=2/3 \)) over the entire voltage range.  At the
lower \Galvani{} limit, the electrolyte is essentially pure
\Otherion{} (\( \Fraction{\Otherion{}}^{\Electrolyte} = 1 \)).  At the upper 
\Galvani{} limit, the electrolyte is pure \Cation{}\Anion{} (\(
\Fraction{\Cation{+2}}^{\Electrolyte} = 1/2 \), \(
\Fraction{\Anion{-2}}^{\Electrolyte} = 1/2 \)).

\subsection{Other Parameters}
\label{sec:Parameters:Others}

A simplification, which should be eliminated in future work, is employed for the
barrier height \Barrier{j}.  We set equal values for the substitutional species
\( \Barrier{j\in 2\ldots\Components} = \Barrier{} \), and the value for the
electrons, \( \Barrier{\Electron{}} = 0 \).  This makes the last term in
Eq.~\eqref{eq:Partial:Phase} independent of \Concentration{j} and eliminates the
\Barrier{j} dependence in Eq.~\eqref{eq:Governing:Electrochemical}.  

In phase
field models of solidification, where electrostatic effects are not included, the
surface \PRE{free} energy \( \SurfaceEnergy_\Phase \) and the interfacial thickness \(
\Thickness_\Phase \) (of the \Phase -field) are given by \cite{Wheeler:1992}
\begin{align}
    \SurfaceEnergy_\Phase 
    &= \sqrt{\frac{\Gradient{\Phase}\Barrier{}}
	{18\PartialMolarVolume{\Substitutional}}}
    \label{eq:Phase:SurfaceEnergy} \\
    \Thickness_\Phase 
    &= \sqrt{\frac{\Gradient{\Phase}\PartialMolarVolume{\Substitutional}}
	{2\Barrier{}}}.
    \label{eq:Phase:Thickness}
\end{align}
We include the subscript \Phase\ with the realization that there will be an
electrostatic contribution to the surface \PRE{free} energy (see
Eq.~\eqref{eq:SurfaceEnergy}) and an independent electrostatic length scale.  For
numerical calculations, we choose the approximate values of \(
\SurfaceEnergy_\Phase = \unit{0.2}{\joule\per\meter\squared} \) and \(
\Thickness_\Phase = \unit{\Scientific{3}{-11}}{\meter} \), which give \(
\Barrier{} = \unit{\Scientific{3.6}{5}}{\joule\per\mole} \) and
\( \Gradient{\Phase} = \unit{\Scientific{3.6}{-11}}{\joule\per\meter} \).  \PRE{In reality, we would not
expect a metallic electrode to have the same permittivity as an aqueous
electrolyte.  Moreover, the permittivity of the electrolyte is known to be lower
near the interface than in the bulk as a result of the polarization in the double
layer} \cite{BockrisReddy:2nd,Goodisman:1987}.  \PRE{While the variation in the
permittivity undoubtedly affects the structure of the interface, our goal in this
paper is to show the richness obtained from a phase field model with even the
simplest assumptions.  We defer examination of phase and concentration dependence
of the permittivity to future work,} and take \( \Dielectric\left(\Phase\right) =
78.49 \PermittivityVacuum \), where \PermittivityVacuum\ is the permittivity of
free space.  This is the value typically cited for an aqueous electrolyte
\cite{Grahame:1947}.

\section{Numerical Methods}
\label{sec:Methods}

Numerical solutions to the governing equations
Eq.~\eqref{eq:Governing} were obtained by both a relaxation method and
a pseudo-spectral technique on a one-dimensional domain of length 
\BoxSize.  The relaxation method had the advantages
that it was simple to code and would eventually converge to a solution
even from a step-function initial condition.  The main disadvantage of
the technique is that it is very slow.  Run times of several hours to
several days were needed to reach convergence on a
\unit{1.6}{\giga\hertz} AMD Athlon running Debian GNU/Linux v3.0 with
a 2.4 kernel and using the Portland Group \texttt{pgcc}
compiler.\footnote{Certain commercial products are identified in this
paper in order to adequately specify procedures being described.  In
no case does such identification imply recommendation or endorsement
by the National Institute of Standards and Technology, nor does it
imply the material identified is necessarily the best for the
purpose.} In contrast, the pseudo-spectral technique can produce
solutions in a few minutes, but only from a very good initial guess.

\subsection{Relaxation}
\label{sec:Methods:Relaxation} 

Relaxation solutions \cite{NumRec:C} 
to Eq.~\eqref{eq:Governing} were
obtained by casting the equilibrium ordinary differential equations
\eqref{eq:Governing:Electrochemical} and \eqref{eq:Governing:Phase} as the time
dependent partial differential equations 
\begin{subequations}
\begin{align}
    \frac{\partial\Concentration{j}}{\partial\Time}
    &= \nabla\cdot
    \left\{
        \Mobility{j}\nabla \frac{\Variation \Lagrangian}{\Variation \Concentration{j}}
    \right\}
    = \nabla\cdot\left\{
	\Mobility{j}\nabla\left[
	    \Electrochemical{j} 
	    - \frac{\PartialMolarVolume{j}}
		{\PartialMolarVolume{\Solvent{}}}
	    \Electrochemical{\Solvent{}}
	\right]
    \right\} \\
    & \qquad\qquad\qquad j = 1, \ldots, \Components-1
    \nonumber
    \label{eq:Evolution:Diffusion}
\end{align}
and
\begin{equation}
    \frac{\partial\Phase}{\partial\Time} 
    = -\Mobility{\Phase} \frac{\Variation \Lagrangian}{\Variation \Phase}
    = -\Mobility{\Phase}\left[
	\frac{\partial\HelmholtzPerVol}{\partial\Phase}
	- \Gradient{\Phase}\nabla^2\Phase
	- \frac{\Dielectric'(\Phase)}{2} 
	    \left(\nabla\Potential\right)^{2}
    \right],
    \label{eq:Evolution:Phase}
\end{equation}%
    \label{eq:Evolution}%
\end{subequations}%
where \Time\ is time, \Mobility{j} is the mobility of component \( j \), and
\Mobility{\Phase} is the mobility of the phase field.  We defer discussion of the
mobilities \Mobility{\Phase} and \Mobility{j} to our paper on kinetics
\cite{ElPhF:kinetics} as their values are not important to the present analysis
of electrochemical equilibrium.  Equations~\eqref{eq:Evolution} are the simplest
expressions that guarantee a decrease in total free energy with time.  Poisson's
equation \eqref{eq:Governing:Poisson} must still be satisfied everywhere.
Equations \eqref{eq:Governing:Poisson} and \eqref{eq:Evolution} were solved with
explicit finite differences.  Spatial derivatives were taken to second order on a
uniform mesh.  Solutions were integrated to equilibrium with an adaptive,
fifth-order Runge-Kutta time stepper
\cite{NumRec:C}. 
We defined equilibrium as the point when
Eqs.~\eqref{eq:Governing:Electrochemical} and
\eqref{eq:FirstIntegral:Equilibrium} were satisfied to within 0.1\%.

Simulations were started with an abrupt interface between the bulk electrode and
electrolyte phases, such that \( \Phase^{\Electrode} = 1 \) and \(
\Phase^{\Electrolyte} = 0 \).  After choosing a value for \(
\Concentration{\Cation{+}}^{\Electrolyte} \), the remaining bulk
\Concentration{j} were determined from Figure~\ref{fig:Nernst}.  Because \(
\ChargeDensity = 0 \) for the bulk concentrations,
Eq.~\eqref{eq:Governing:Poisson} gives \( \nabla\Potential = 0 \) throughout the
domain for the initial data.  The boundary conditions are listed in
Table~\ref{tab:BoundaryConditions}.
\begin{table}[tbp]
    \centering
    \caption{Boundary Conditions}
    \begin{ruledtabular}
	\begin{tabular}{==}
	    \multicolumn{1}{c}{\text{electrode (\( \Position = 0 \))}} 
	    & \multicolumn{1}{c}{\text{electrolyte (\( \Position = \BoxSize \))}}
	    \\
	    \hline
	    
	    \vec{\Normal}\cdot\nabla\Phase = 0 
	    & \vec{\Normal}\cdot\nabla\Phase = 0
	    \\
	    \Potential = 0 
	    & \vec{\Normal}\cdot\nabla\Potential = 0
	    \\
	    \multicolumn{1}{|}{\Concentration{j} | \text{specified}}
	    & \multicolumn{1}{|}{\Concentration{j} | \text{specified}}
	\end{tabular}
    \end{ruledtabular}
    \label{tab:BoundaryConditions}
\end{table}

\subsection{Adaptive pseudospectral discretization}
\label{sec:Methods:Spectral} 

In order to increase the numerical resolution of the interfacial
region, we have also employed an adaptive solution technique based on
a spectral approximation \cite{VoGH84} to the governing equations
\eqref{eq:Governing}.  To reduce the number of unknowns in the system,
we eliminate the solute variables by solving the governing equations
\eqref{eq:Governing:Electrochemical}, together with the constraint
equation \eqref{eq:VolumeConstraint}.  Given values of \Phase\ and
\Potential\ at a point, this provides an algebraic form for the solute
fields \Concentration{j} at this point.  The remaining equations
\eqref{eq:Governing:Phase} and \eqref{eq:Governing:Poisson} are
discretized using a pseudospectral formulation of a spectral element
representation of the second derivative \cite{Pate84}.  It is
convenient to fix the interface location by specifying
\(\Phase(\Position_I) = 1/2\) at a given grid point \(
\Position_I \).  The discretization procedure then provides a set of
nonlinear equations with an equal number of unknowns.  The nonlinear
equations are solved using the quasi-Newton software package \textsc{SNSQ}
\cite{SNSQ}.  Starting estimates for the solution procedure are
generally obtained by continuation from the finite difference
procedure described above, or from previous pseudospectral solutions.

An adaptive procedure is obtained by bisecting the elements for which
an error estimate indicates that additional refinement is necessary. 
The error estimate is based on the rate of decay of a Chebyshev
expansion of the solution components; a simple criterion is based on
requiring the magnitude of the last two coefficients of the charge
density in each panel to lie below a given threshold.  If a refinement
is necessary, the element is bisected and the previous solution is
interpolated to the nodes of the new panels.  The nonlinear equations
are then solved on the new nodes, and the procedure is repeated until
each Chebyshev expansion has rapid decay, indicating that the solution
is well-resolved on each panel.  Since the previous solution provides
a good starting guess, the successive solutions converge quickly, and
the overall run time is a small multiple of the time required for a
single solution step on a grid with equal pseudospectral elements.

\section{Numerical Results}
\label{sec:Solutions}

\begin{figure}[tbp]
    \centering
    \subfigure{
	\includegraphics[width=\figurewidth]{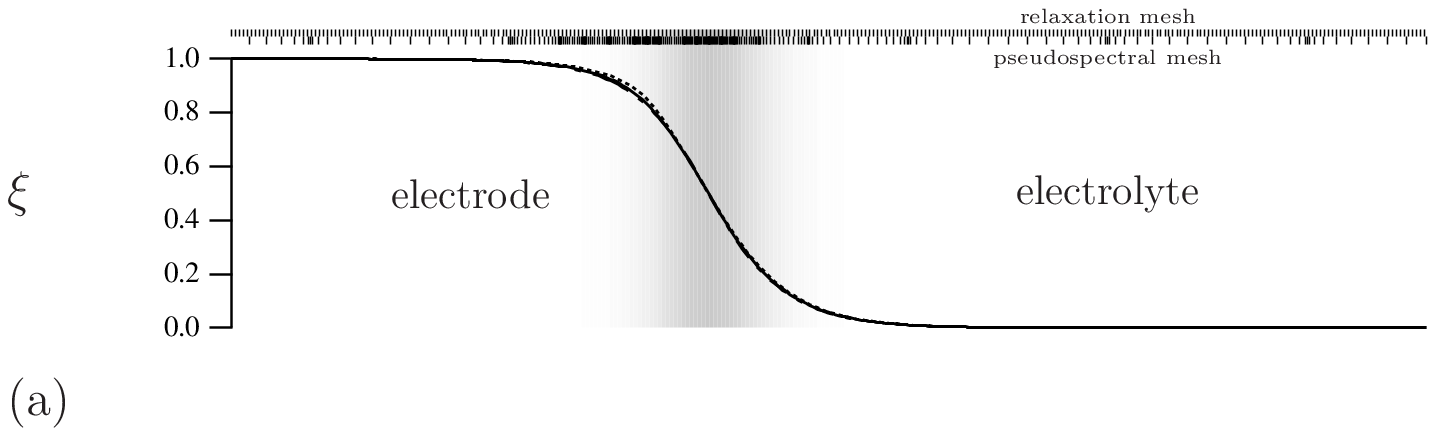}
	\protect\label{fig:PhaseField}
    }
    \subfigure{
	\includegraphics[width=\figurewidth]{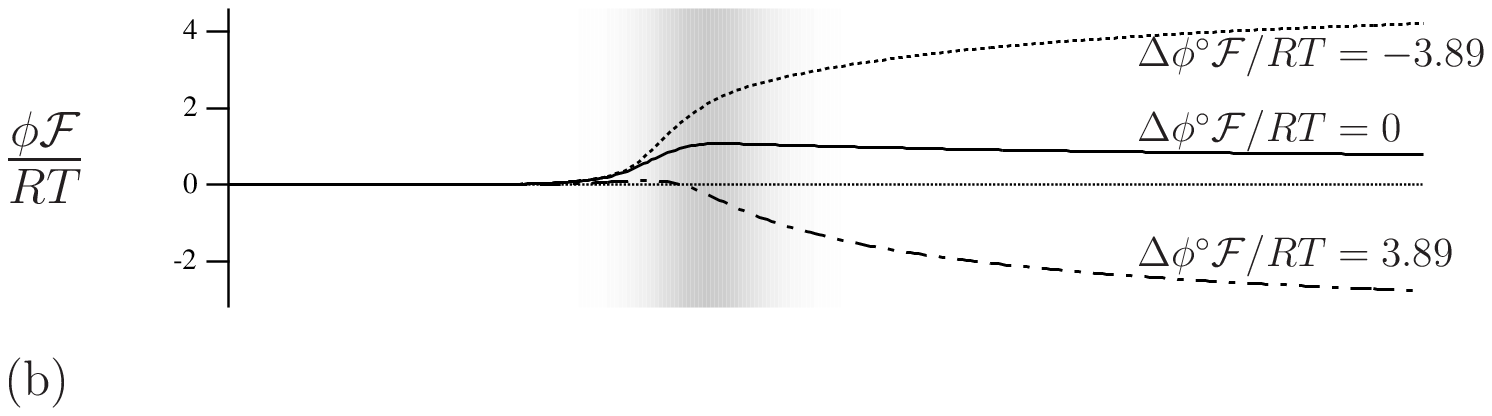}
	\protect\label{fig:Voltage}
    }
    \subfigure{
	\includegraphics[width=\figurewidth]{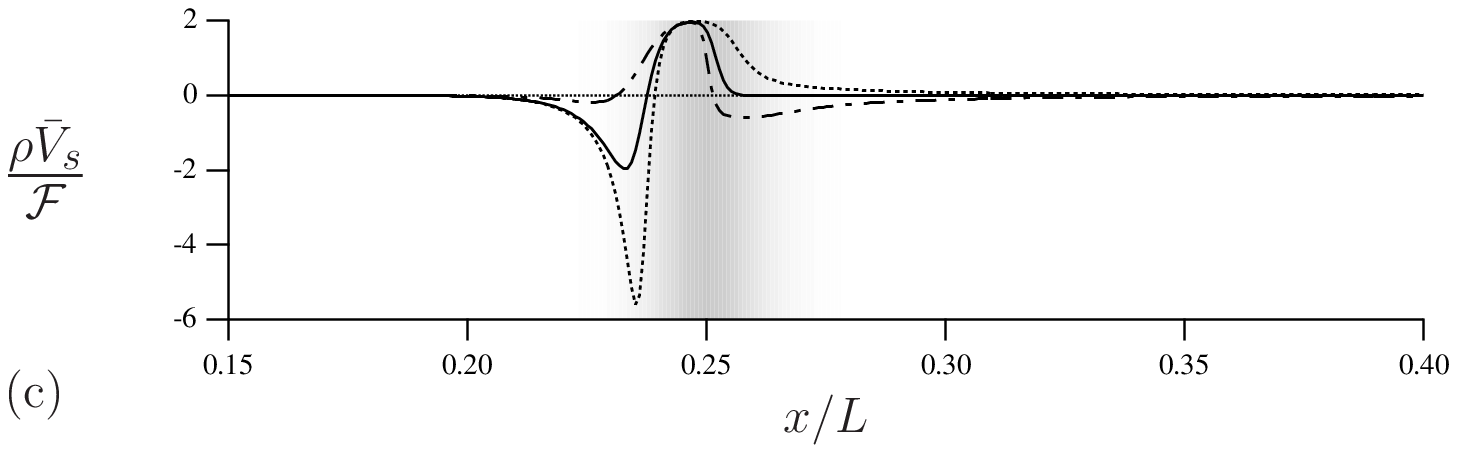}
	\protect\label{fig:Charge}
    } 
    \caption{Profiles through the interface \PRE{of (a) the phase field
    variable, (b) the normalized electrostatic potential, and (c) the normalized
    charge distribution} for \( \Galvani{\Standard} = \unit{-0.1}{\volt} \),
    \unit{0}{\volt}, and \unit{+0.1}{\volt}.  \( \DoubleWell(\Phase) \) is mapped
    onto the background in gray to indicate the location of the \PRE{phase field}
    interface.  The ticks marks at the top of Figure~(a) indicate the positions
    of the mesh points for the two different solution methods.}
    \protect\label{fig:Profiles}
\end{figure}
Figure~\ref{fig:Profiles} shows plots of phase field, voltage and charge across
the interface for \( \Galvani{\Standard} = \unit{(-0.1, 0, +0.1)}{\volt} \)
obtained using the relaxation method in 1D over a domain \unit{3.2}{\nano\meter}
long and containing 1200 mesh points.  The tick marks at the top of
Figure~\ref{fig:PhaseField} indicate the positions of the mesh points used in the
two different solution methods.  \PRE{The fields calculated by the relaxation and
the pseudospectral methods are indistinguishable on the scale of these graphs, so
we apply a linear correlation function to compare the calculation methods.} The
two methods have a linear correlation of 0.9992 for the most sensitive field
\ChargeDensity; the other fields have a linear correlation of 0.9999 or better.
\PRE{The difference between the two methods for the most sensitive field is thus
of the same order as the criterion for stopping the relaxation calculations
(\(<0.1\%\) error in Eqs.~\eqref{eq:Governing:Electrochemical} and
\eqref{eq:FirstIntegral:Equilibrium}); all other fields are much closer to
convergence.}

Figure~\ref{fig:Profiles} focuses on the interface region of this computational
domain.  The bulk concentration of \Cation{+} and \Anion{-} in the electrolyte is
\unit{0.25}{\mole\per\liter}.  The variation of \Phase\ between the electrode on
the left and the electrolyte on the right for the three cases is virtually
identical.  A fit of all three curves to \( \Phase(\Position) = \{1-\tanh
[(\Position - \Position_{0}) / 2\Thickness_{\Phase}] \}/2 \) gives \(
\Thickness_{\Phase} = \unit{\Scientific{(2.480 \pm 0.009)}{-11}}{\meter} \),
which compares well with the value we assumed in
Section~\ref{sec:Parameters:Others}.  \Phase\ changes from 0.9 to 0.1 over a
distance of approximately \unit{0.1}{\nano\meter} or \( 4 \Thickness_\Phase \).
This represents the thickness of the electrode-electrolyte interface.
The voltage \Potential\ changes smoothly between a value of zero in the electrode
(as assumed) to an asymptotic value in the electrolyte far from the interface
equal to \( (\Gas\Temperature/2\Faraday)\ln
[(\unit{1}{\mole\per\liter})/(\unit{0.25}{\mole\per\liter})] -
\Galvani{\Standard} \).  The charge \ChargeDensity, while zero far from the
interfacial region, exhibits a distinct charge separation within the interfacial
region.  The charge distribution is quite different for the three cases.  For \(
\Galvani{\Standard} = \unit{-0.1}{\volt} \) and \unit{0.0}{\volt}, a negative
charge is present to the left.  For \( \Galvani{\Standard} = \unit{+0.1}{\volt}
\), the negative charge is to the right.
\begin{figure}[tbp]
    \centering
    \subfigure{
	\includegraphics[width=\figurewidth]{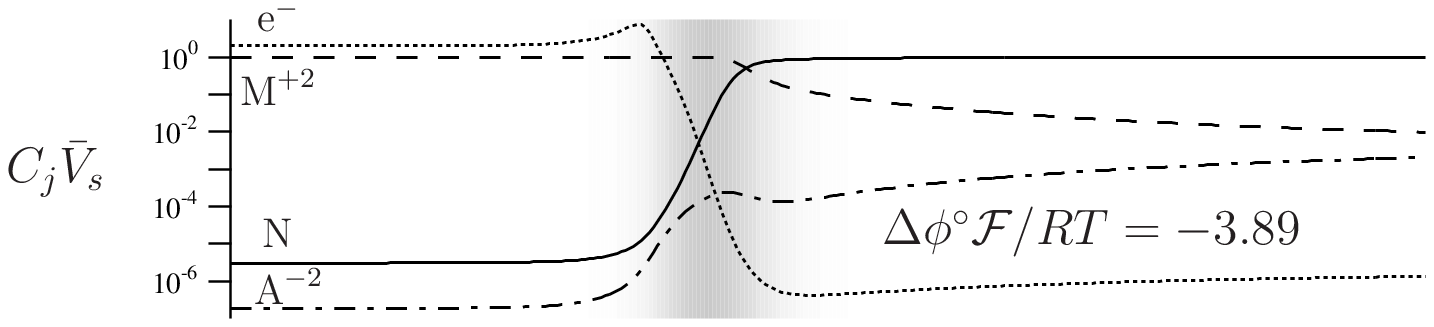}
	\protect\label{fig:Concentration:Minus}
    }
    \subfigure{
	\includegraphics[width=\figurewidth]{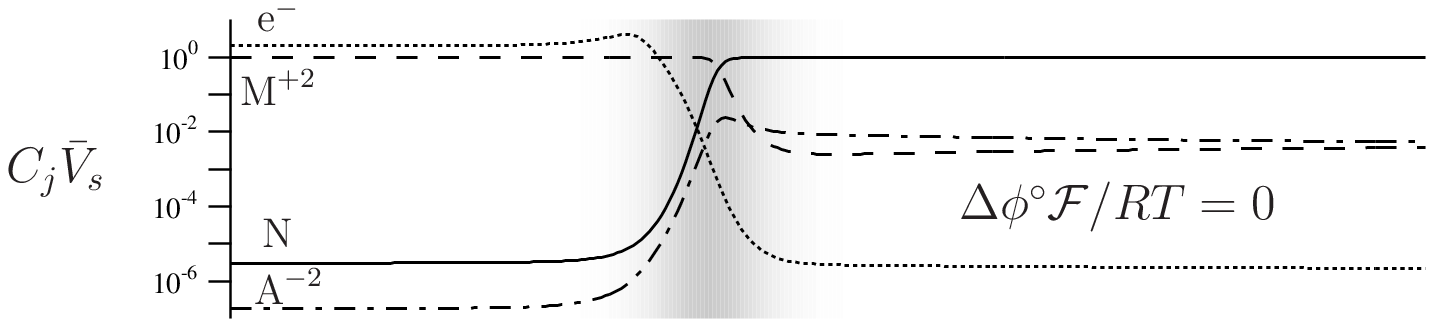}
	\protect\label{fig:Concentration:Zero}
    }
    \subfigure{
	\includegraphics[width=\figurewidth]{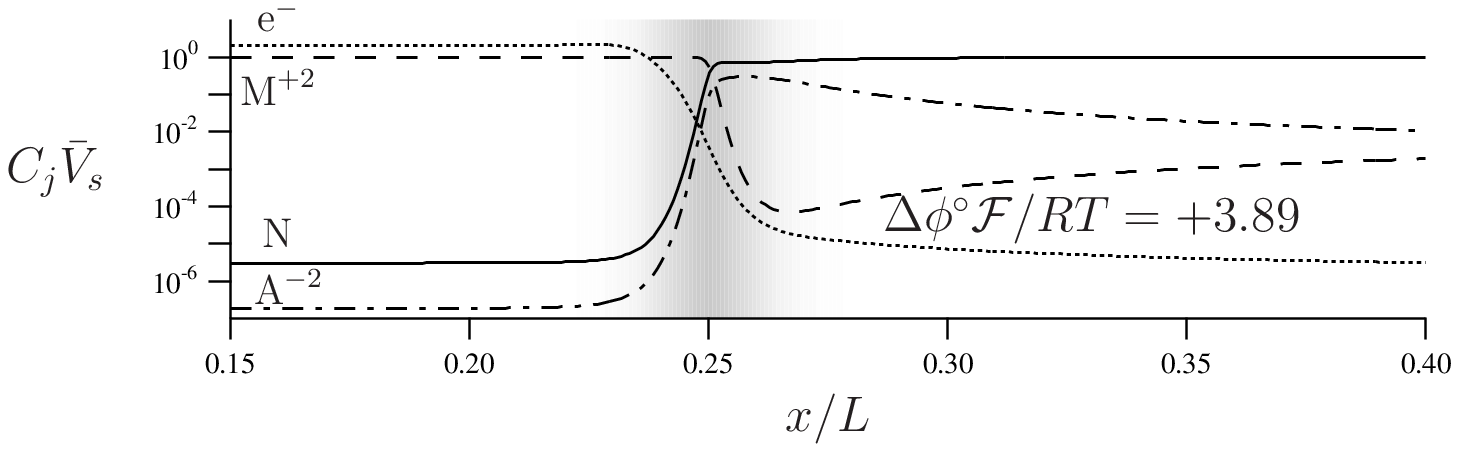}
	\protect\label{fig:Concentration:Plus}
    } 
    \caption{\PRE{Normalized} concentration profiles through the interface for
    different values of \Galvani{\Standard}.  \( \DoubleWell(\Phase) \) is mapped
    onto the background in gray to indicate the location of the \PRE{phase field}
    interface.}
    \protect\label{fig:Concentration}
\end{figure}
Figure~\ref{fig:Concentration} shows the variation in the concentrations from
the electrode to the electrolyte.  The values at the ends of the full
computational domain correspond to a bulk \Cation{} electrode, a
\unit{0.25}{\mole\per\liter} \Cation{}\Anion{} in \Otherion{} electrolyte, with
impurities as allowed by Table~\ref{tab:Parameters:Thermodynamic:Aqueous}.  The
abrupt change in concentrations through the distance where \Phase\ is changing
is followed by a more gradual change in the electrolyte.  The gradual
concentration decay length in the electrolyte is the same as that of the
voltage. One could define the surface excess as the difference between the 
actual concentration and some interpolation between the bulk values and see that 
there is an adsorption of the different species at the interface which depends on 
the value of \Galvani{\Standard}.

In Appendix~\ref{sec:Classical:GouyChapmanStern} we summarize the
Gouy-Chapman-Stern model of the double layer.  That treatment predicts an
exponential decay of the potential in the electrolyte away from the electrode,
with a decay length of \( \Thickness_{\Potential}^\text{GC} \).
\begin{figure}[tbp]
    \centering
    \includegraphics[width=\figurewidth]{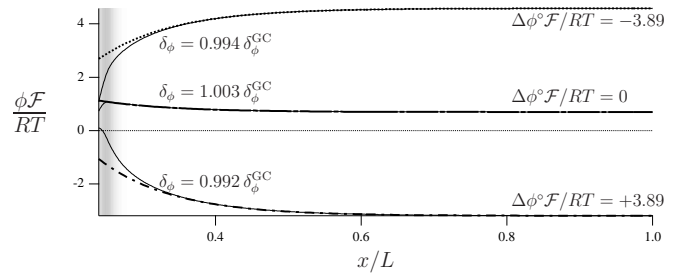}
    \caption{Exponential fits (heavy dashed lines) to potential curves of
    Figure~\ref{fig:Voltage} (light solid lines).  \( \DoubleWell(\Phase) \) is
    mapped onto the background in gray to indicate the location of the
    \PRE{phase field} interface.}
    \protect\label{fig:Voltage:Fit}
\end{figure}
Figure~\ref{fig:Voltage:Fit} shows a fit of the \Potential\ vs.\
distance plots from Figure~\ref{fig:Voltage} to \( \Potential =
\Potential_{\infty} + (\Potential_{\IHP} - \Potential_{\infty}) \exp
(-\Position / \Thickness_{\Potential}) \).  The fit is excellent.  The
decay length \( \Thickness_{\Potential} \) of \Potential\ to its
asymptotic value is very close to the predicted value of \(
\Thickness_{\Potential}^\text{GC} \).  This length is over ten times
larger than \( \Thickness_{\Phase} \) and approximately three times
the apparent interface thickness.

\begin{figure}[tbp]
    \centering
    \subfigure{\label{fig:SurfaceEnergy}%
	\includegraphics[width=\subfigurewidth]{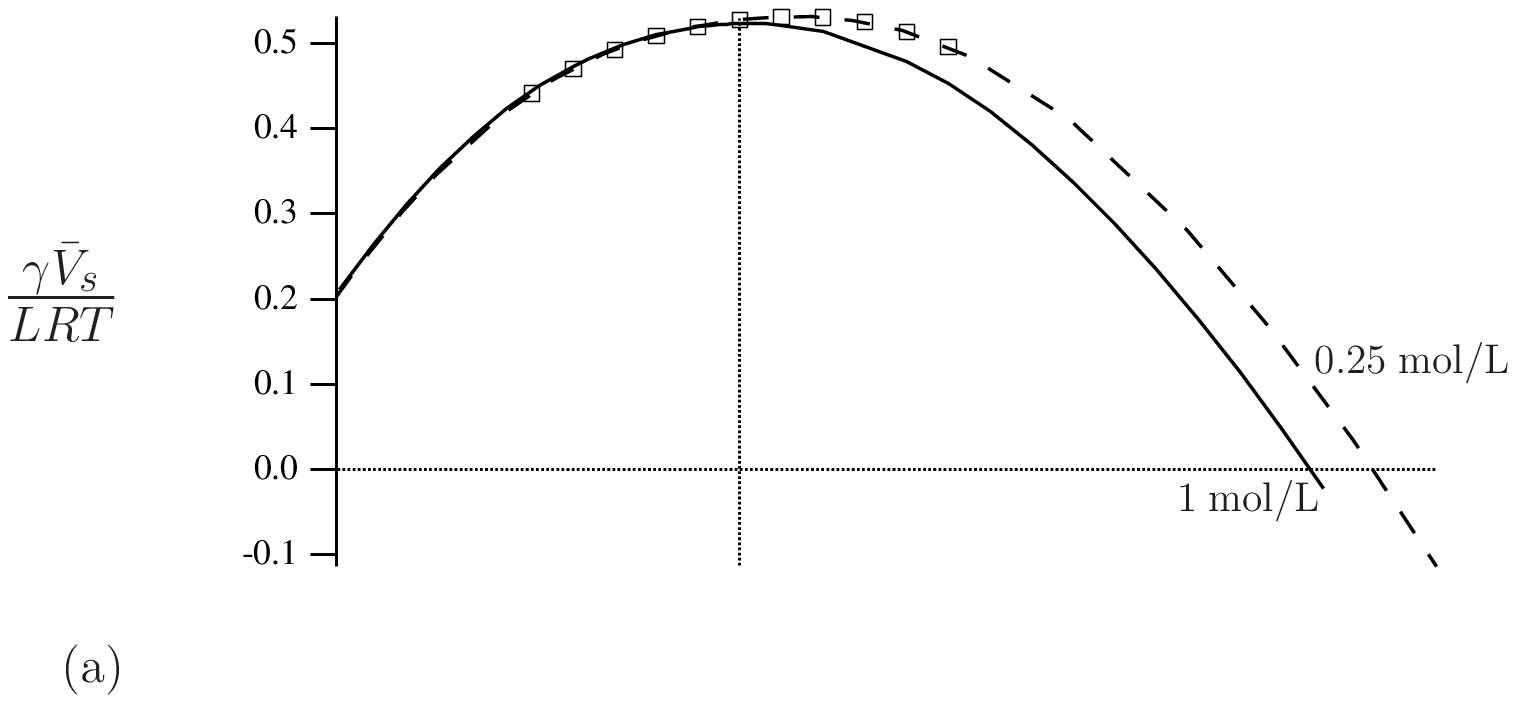}}\\
    \subfigure{\label{fig:SurfaceCharge}%
	\includegraphics[width=\subfigurewidth]{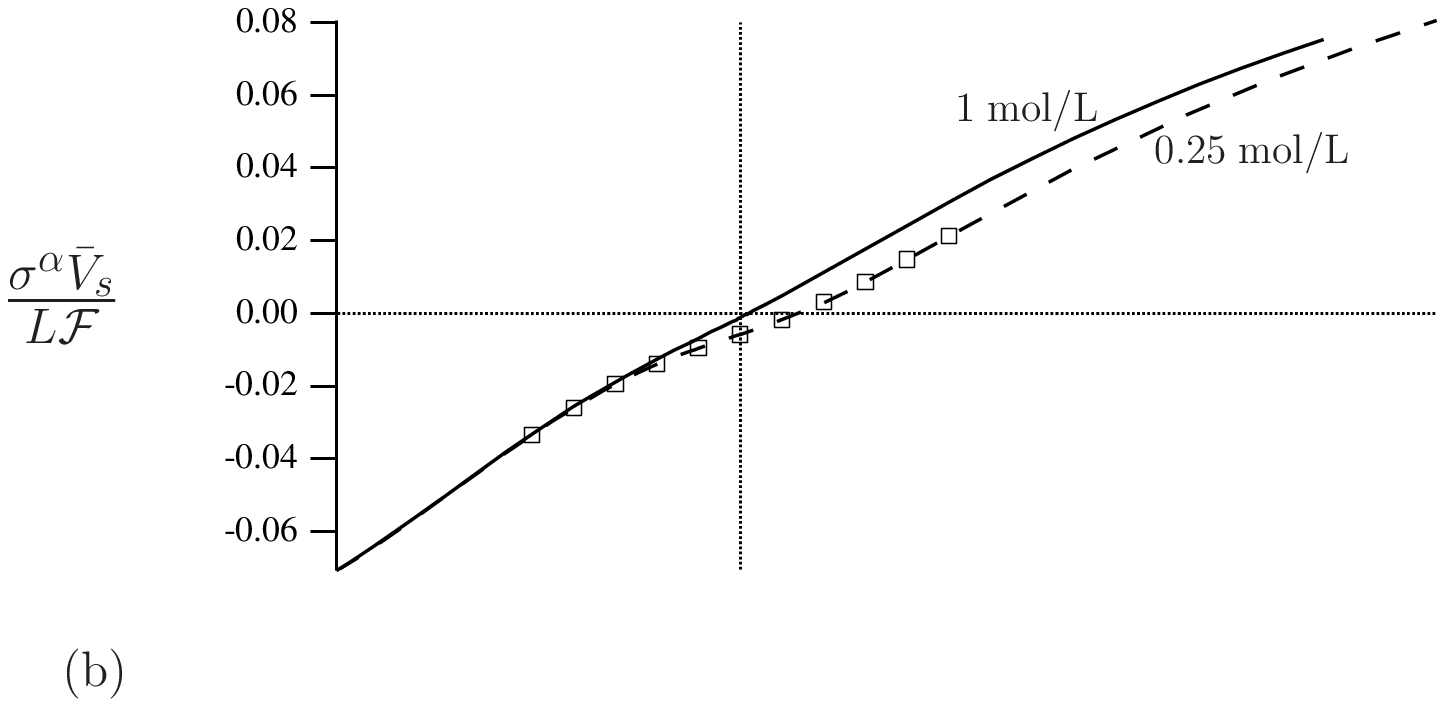}}\\
    \subfigure{\label{fig:DifferentialCapacitance}%
	\includegraphics[width=\subfigurewidth]{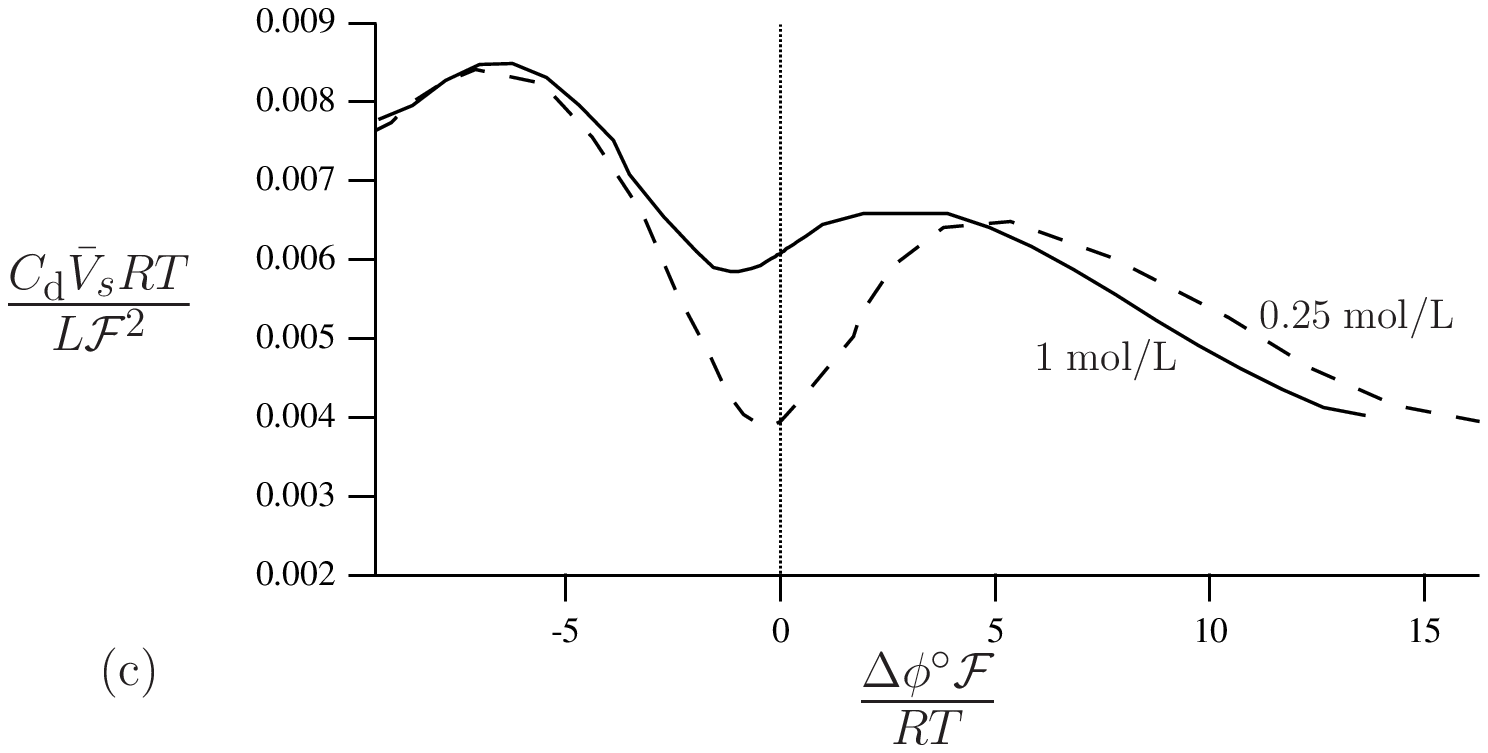}}%
    \caption{\PRE{(a) Normalized surface free energy, (b) normalized surface
    charge, and (c) and normalized differential capacitance} as functions of
    normalized \Galvani{\Standard}.  Open symbols are calculated by the
    relaxation method of Section~\ref{sec:Methods:Relaxation}.  Lines are
    calculated by the pseudospectral method of
    Section~\ref{sec:Methods:Spectral}.}
    \label{fig:Spectral}
\end{figure}
Figure~\ref{fig:SurfaceEnergy} shows the surface \PRE{free} energy (from
Eq.~\eqref{eq:SurfaceEnergy}) vs.\ \Galvani{\Standard} and
Figure~\ref{fig:SurfaceCharge} shows a plot of \(
\SurfaceCharge^\Electrode \) vs.  \Galvani{\Standard}, both obtained
by the two numerical methods.  The point of zero charge (PZC), defined
by \( \SurfaceCharge^{\Electrode} = \SurfaceCharge^{\Electrolyte} = 0
\), occurs at \( \Galvani{\Standard} = \unit{+0.005}{\volt} \) for
\unit{1}{\mole\per\liter} and \( \Galvani{\Standard} =
\unit{+0.035}{\volt} \) for \unit{0.25}{\mole\per\liter}.  At the
point of zero charge, \ChargeDensity\ is not constant, nor is the
electrostatic potential, but rather, the integrated charge is zero in
each phase and there is some potential step between them.  We note
that Grahame described exactly this condition in his seminal paper on
the electrochemical double-layer \cite{Grahame:1947}.  The presence of
dipoles at the interface guarantees that the potential will
not be uniform.  The surface charge curve shows a slight deviation
from linearity away from the point of zero charge.  This dome shaped
curve in Figure~\ref{fig:SurfaceEnergy} has a maximum surface \PRE{free} energy
of approximately \unit{0.225}{\joule\per\meter\squared} at a value of
\( \Galvani{\Standard} = \unit{+0.005}{\volt} \), the point of zero
charge.  This maximum surface \PRE{free} energy value is very close to \(
\SurfaceEnergy_\Phase \), used to establish numerical values for
\Barrier{} and \( \Gradient{\Phase} \). 
Figures~\ref{fig:SurfaceEnergy} and \ref{fig:SurfaceCharge} obey
Eq.~\eqref{eq:Adsorption:Classical} very closely.  The negative
surface \PRE{free} energies obtained for large positive values of
\Galvani{\Standard} indicate that a planar interface will become unstable
to perturbations which increase surface area.  Such perturbations
are not possible given the symmetry constraints of our 1D solutions,
but attention will need to be paid to this when higher dimensional
calculations are performed.

From Eq.~\eqref{eq:DifferentialCapacitance}, the differential capacitance is
obtained as the derivative of Figure~\ref{fig:SurfaceCharge} with respect to
\Galvani{\Standard}.  The relaxation method used to produce the open square
points in Figure~\ref{fig:SurfaceCharge} is not fast to enough to allow
calculating a numeric derivative of sufficient resolution.  We thus use the
results of the spectral method, which can compute with a much greater resolution
and over a wider range of \Galvani{\Standard}, to calculate
Figure~\ref{fig:DifferentialCapacitance}.

\begin{figure*}[tbp]
    \centering
    \subfigure{\label{fig:Capacitance:ElPhF}%
        \includegraphics[width=0.45\textwidth]{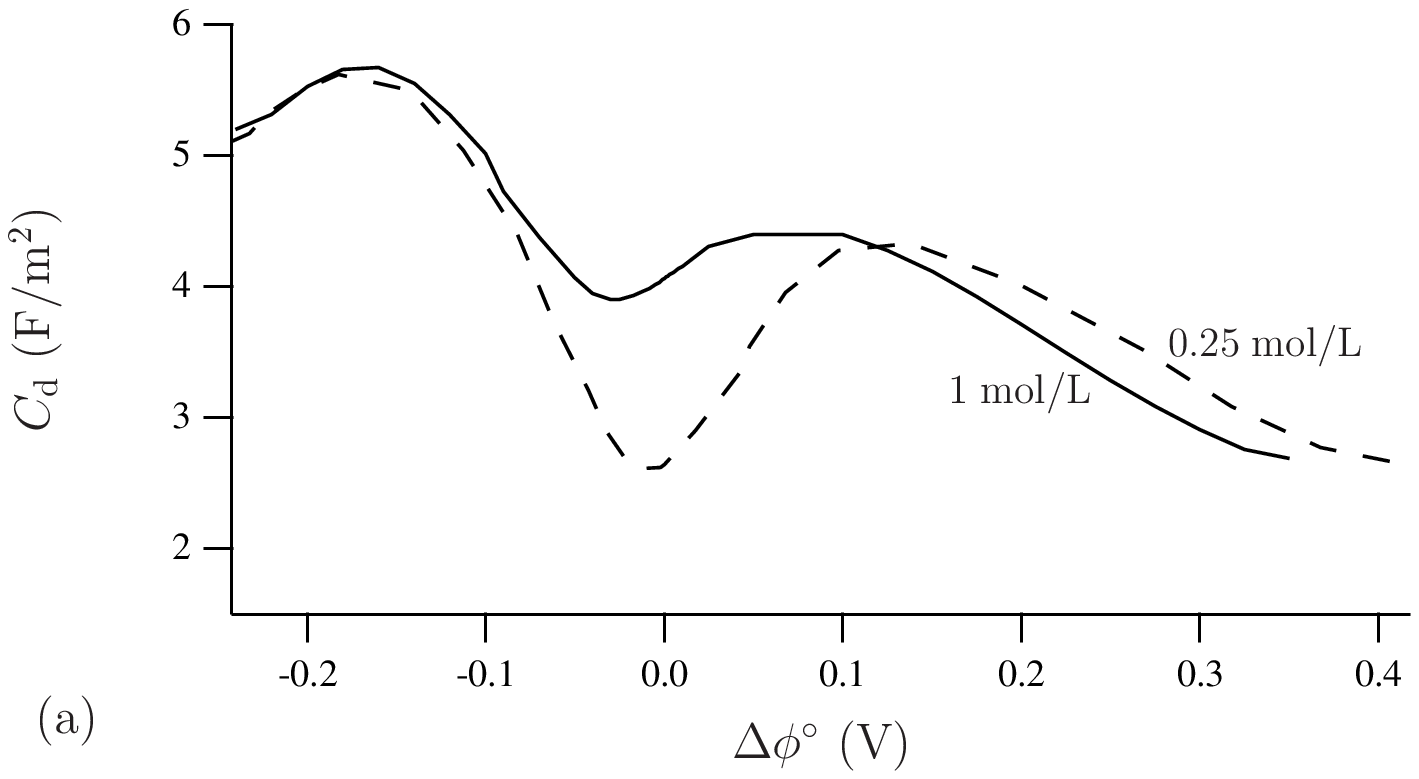}}
    \subfigure{\label{fig:Capacitance:GC}%
        \includegraphics[width=0.45\textwidth]{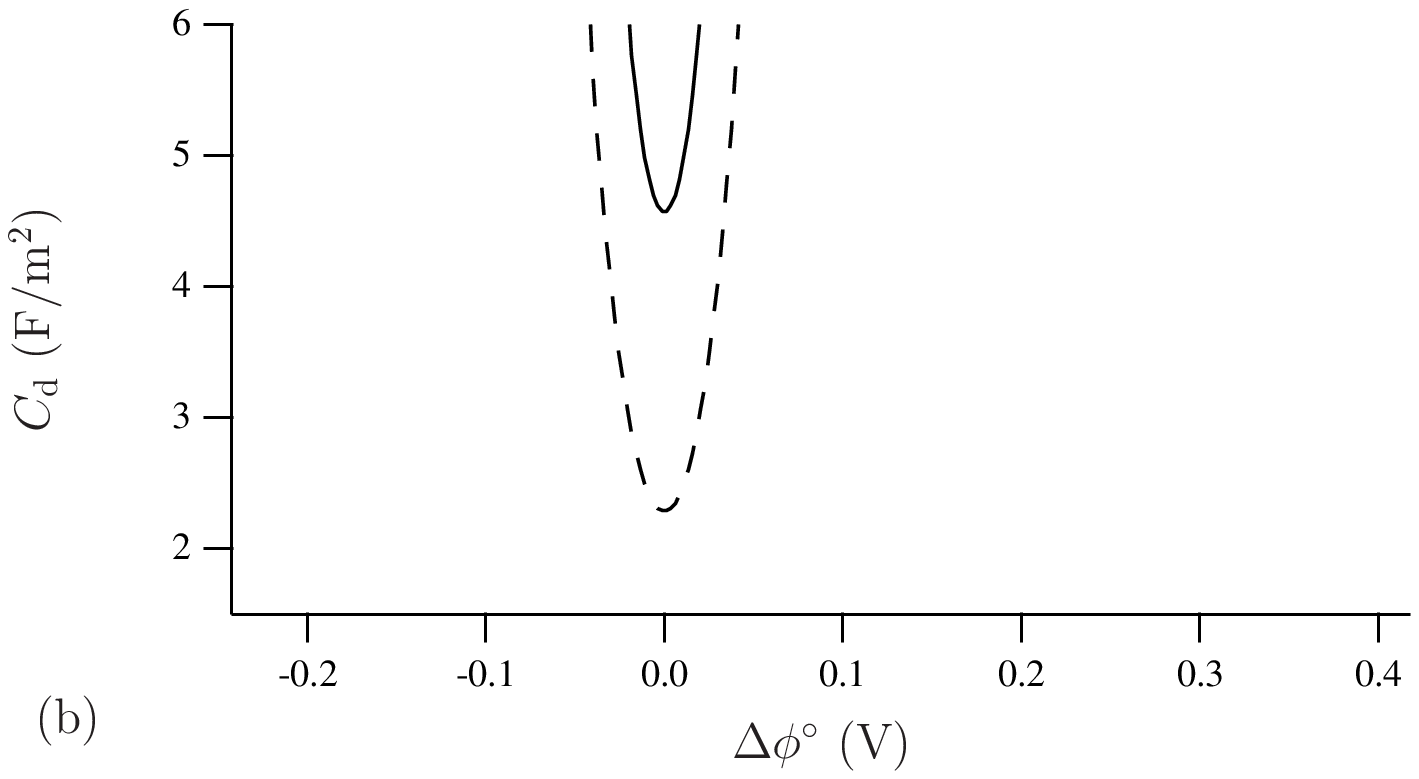}}\\
    \subfigure{\label{fig:Capacitance:GCS}%
        \includegraphics[width=0.45\textwidth]{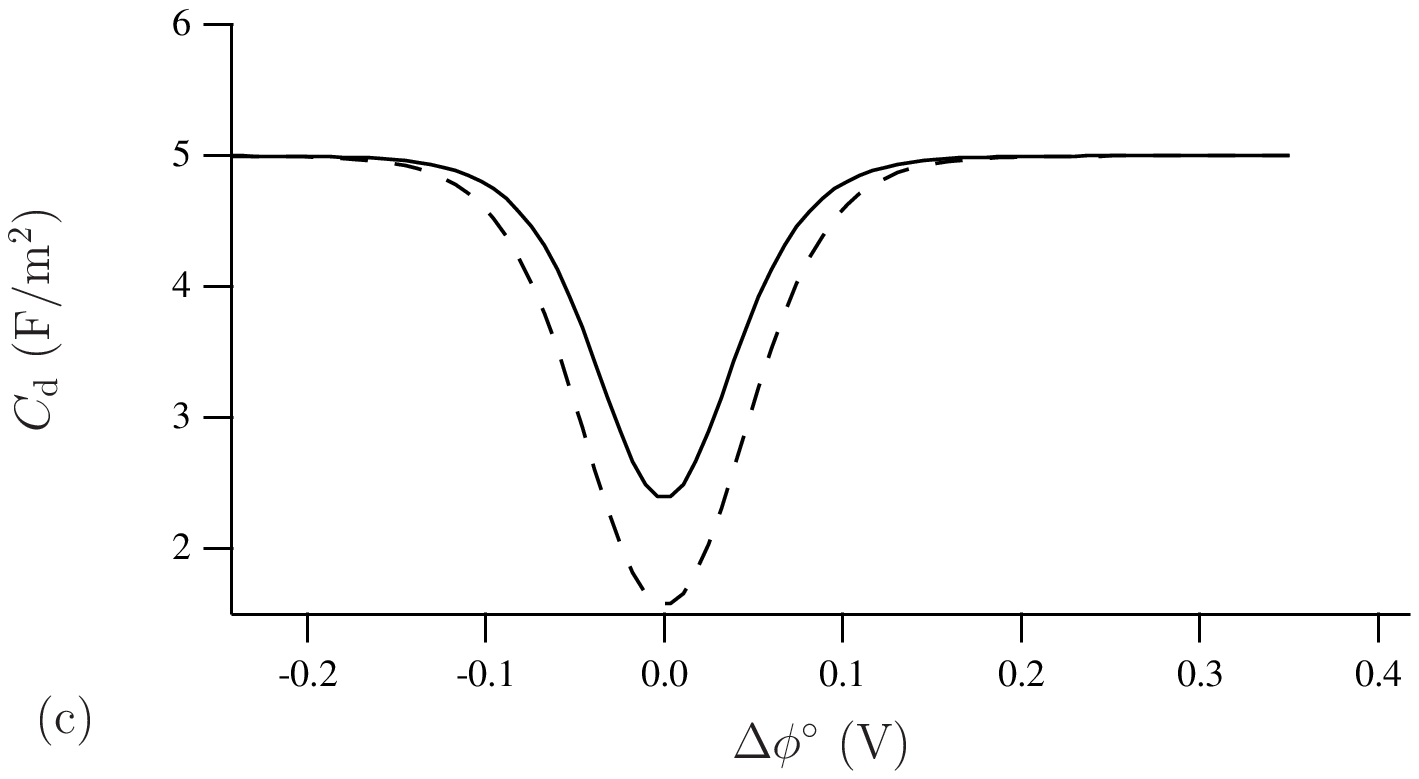}}
    \subfigure{\label{fig:Capacitance:Valette}%
        \includegraphics[width=0.4\textwidth]{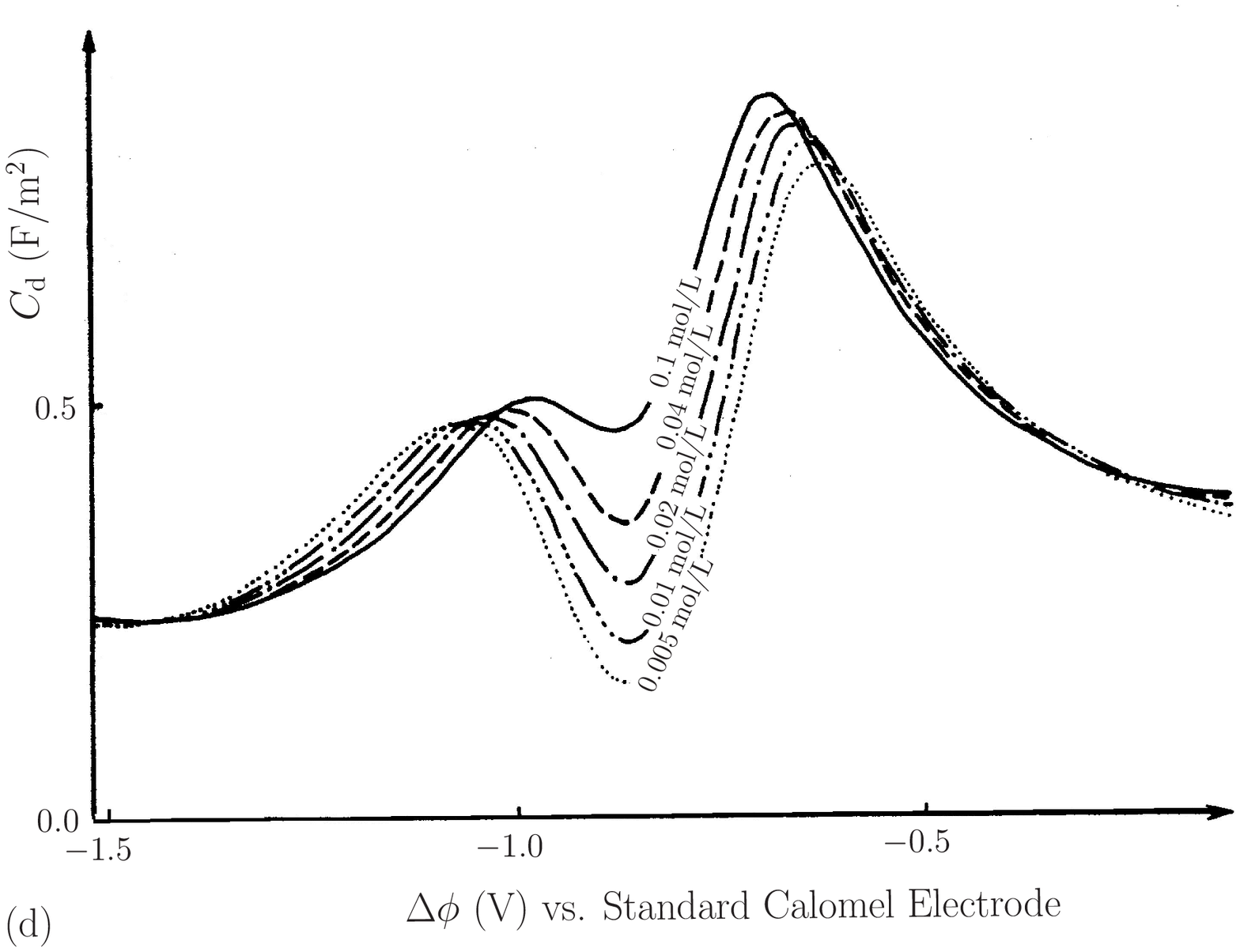}}%
    \caption{Comparison of the differential capacitance results of \PRE{(a)} this
    model with the predictions of \PRE{(b) the Gouy-Chapman and (c)
    Gouy-Chapman-Stern} sharp interface theories outlined in
    Appendix~\protect\ref{sec:Classical:GouyChapmanStern} and \PRE{(d)} the 
    experimental measurements of Ag(100) electrodes in aqueous solutions of NaF
    [Reprinted from \textit{J. Electroanal.  Chem.} \textbf{138}, G. Valette,
    ``Double layer on silver single crystal electrodes in contact with
    electrolytes having anions which are slightly specifically adsorbed Part II.
    The (100) face'', 37--54, Copyright (1982), with permission from Elsevier].}
    \label{fig:Capacitance}
\end{figure*}
Our calculated differential capacitance curve, replotted in
Figure~\ref{fig:Capacitance:ElPhF}, does not resemble the hyperbolic cosine
predicted by the Gouy-Chapman theory (\( \Position_{2} = 0 \) in
Eq.~\eqref{eq:GC:capacitance}), shown in Figure~\ref{fig:Capacitance:GC}.
Neither does it resemble the truncated hyperbolic cosine predicted by the
Gouy-Chapman-Stern theory (\( \Position_{2} \neq 0 \) in
Eq.~\eqref{eq:GC:capacitance}), shown in Figure~\ref{fig:Capacitance:GCS} (we
take \( \Dielectric / \Position_{\OHP} = \unit{5}{\farad\per\squaren\meter}\) for
illustration only).  On the other hand, it does bear a striking resemblance to
experimental differential capacitance curves
\cite{Grahame:1947,Valette:1973,Valette:1982}, such as Valette and Hamelin's
measurements of Ag electrodes in NaF aqueous solutions, shown in
Figure~\ref{fig:Capacitance:Valette}. \PRE{The density functional calculations 
of Tang, Scriven, and Davis \cite{Tang:1992} also exhibit differential 
capacitance curves with multiple inflection points.}

\section{Discussion and Conclusions}

This paper has explored the equilibrium structure of an electrified interface
between two phases consisting of charged components, as described by a phase
field model.  Such a model, being a continuum description, adds only the bare
essentials of the physics and chemistry of electrochemical interfaces: mass and
volume constraints, Poisson's equation, ideal solution thermodynamics in the
bulk, and a simple description of the competing energies in the interface.
Despite this simple description, the model realizes the oft described behavior of
the double layer; namely, the charge separation at the interface and its
dependence on voltage drop (Galvani or inner potential) across the interface.  As
the Galvani potential is varied at constant compositions of the electrode and
electrolyte (constant chemical potentials), the model reproduces the well known
maximum of the surface \PRE{free} energy curve at the point of zero surface charge (PZC).
High precision pseudospectral solutions of the governing equations also deliver
differential capacitance variations with Galvani potential that exhibit much more
complex and realistic behavior than do the simple Gouy-Chapman-Stern models.  The
full range of behavior encompassed by the model must await further research.  For
example, the effect of unequal and/or nonzero barrier heights \Barrier{j} for the
components will surely affect the adsorption and, in turn, the surface \PRE{free} energy and
capacitance curves.

A recent lattice-gas model of an electrochemical system
\cite{Bernard:2001,Bernard:2003} exhibits interfacial structures very similar to
those found in this paper.  That model also demonstrates simple dendrites
during plating, but those lattice-gas papers do not explore the 
electrocapillary behavior discussed in this paper. The similarities of the 
predictions between that discrete model and our continuum approach may permit a 
bridge between atomistic treatments of the electrochemical interface and 
macroscopic descriptions of electroplating.

To model a real electrochemical system with this method, one needs to
match the parameters of the phase field model to the experimentally
determined (or the normally applied) understanding of the particular
electrochemical system.  In addition to kinetic parameters described
in Ref.~\cite{ElPhF:kinetics}, equilibrium solutions require several
pieces of information.  At a minimum one requires:
\begin{enumerate}
    \item a description of the bulk thermodynamics of the electrode and 
    electrolyte, 
    
    \item the dielectric constant of the electrolyte and electrode,

    \item \label{require-thickness} an estimate for the physical thickness of the electrode /
    electrolyte phase interface,

    \item the actual (Galvani) potential across the interface for some
    concentration of co-existing electrolyte and electrode phases, and

    \item \label{require-surfaceEnergyCapacitance} the surface \PRE{free} energy \emph{or}
    the capacitance of the interface for these concentrations.
\end{enumerate}
Although we currently lack an analytical expression for the relation between the
phase field parameters \Gradient{\Phase} and \Barrier{j} and the information in
\ref{require-thickness} and \ref{require-surfaceEnergyCapacitance} above, the
numerical results of the paper show that they are connected.  In the future, an
asymptotic analysis of the governing equations may reveal these relationships
directly.  Finally, the non-ideal solution behavior of the electrolyte, which may
involve complexing of ions, should be addressed.  Concentration would be replaced
by presumably known activity coefficients.

Solution of the governing equations has proved difficult.  The resolution of the
charge through the interfacial region requires many more mesh points than typical
of phase field models of solidification of binary alloys.  This is due to the
more intricate structure of the charge distribution in the interface as compared
to the structure of the phase and concentration fields.  An adaptive solution
method, which concentrates mesh points in this interfacial region, permits
significantly improved calculation speed.

\begin{acknowledgments}
    J.~W.~Cahn, S.~Coriell, A.~Lobkovsky, R.~F.~Sekerka, and D.~Wheeler have been
    very helpful with our formulation and coding of the phase field model.  We
    have learned a significant amount about electrochemistry through discussions
    with U.~Bertocci, E.~Gileadi, T.~P.~Moffat, and G.~R.~Stafford.  This
    manuscript benefited from a critical reading by E.~Garc\'\i a.  Part of this
    research was supported by the Microgravity Research Division of NASA.
\end{acknowledgments}

\bibliography{abbrTitles,electrochemistry,phaseField,diffusion,Jeff}

\begin{thebibliography}{10}
\expandafter\ifx\csname bibnamefont\endcsname\relax
  \def\bibnamefont#1{#1}\fi
\expandafter\ifx\csname bibfnamefont\endcsname\relax
  \def\bibfnamefont#1{#1}\fi
\expandafter\ifx\csname url\endcsname\relax
  \def\url#1{\texttt{#1}}\fi
\expandafter\ifx\csname urlprefix\endcsname\relax\def\urlprefix{URL }\fi
\expandafter\ifx\csname bibinfo\endcsname\relax \def\bibinfo#1#2{#2}\fi
\expandafter\ifx\csname eprint\endcsname\relax \def\eprint#1{#1}\fi

\bibitem{Josell:2001}
\bibinfo{author}{\bibfnamefont{D.}~\bibnamefont{Josell}},
  \bibinfo{author}{\bibfnamefont{D.}~\bibnamefont{Wheeler}}, \bibnamefont{and}
  \bibinfo{author}{\bibfnamefont{W.~H.} \bibnamefont{Huber}},
  \bibinfo{journal}{Phys. Rev. Lett.} \textbf{\bibinfo{volume}{87}},
  \bibinfo{pages}{016102} (\bibinfo{year}{2001}).

\bibitem{Grahame:1947}
\bibinfo{author}{\bibfnamefont{D.~C.} \bibnamefont{Grahame}},
  \bibinfo{journal}{Chem. Rev.} \textbf{\bibinfo{volume}{41}},
  \bibinfo{pages}{441} (\bibinfo{year}{1947}).

\bibitem{Vetter:1967}
\bibinfo{author}{\bibfnamefont{K.~J.} \bibnamefont{Vetter}},
  \emph{\bibinfo{title}{Electrochemical Kinetics: Theoretical and Experimental
  Aspects}} (\bibinfo{publisher}{Academic Press Inc., New York},
  \bibinfo{year}{1967}).

\bibitem{BockrisReddy:2nd}
\bibinfo{author}{\bibfnamefont{J.~O.} \bibnamefont{Bockris}},
  \bibinfo{author}{\bibfnamefont{A.~K.~N.} \bibnamefont{Reddy}},
  \bibnamefont{and}
  \bibinfo{author}{\bibfnamefont{M.}~\bibnamefont{Gamboa-Aldeco}}, in
  \emph{\bibinfo{booktitle}{Modern Electrochemistry}}
  (\bibinfo{publisher}{Kluwer Academic/Plenum Publishers, New York},
  \bibinfo{year}{2000}), vol.~\bibinfo{volume}{2A}.

\bibitem{Bard:2nd}
\bibinfo{author}{\bibfnamefont{A.~J.} \bibnamefont{Bard}} \bibnamefont{and}
  \bibinfo{author}{\bibfnamefont{L.~R.} \bibnamefont{Faulkner}},
  \emph{\bibinfo{title}{Electrochemical Methods: Fundamentals and
  Applications}} (\bibinfo{publisher}{John Wiley \& Sons, Inc.},
  \bibinfo{address}{New York}, \bibinfo{year}{2001}).

\bibitem{Goodisman:1987}
\bibinfo{author}{\bibfnamefont{J.}~\bibnamefont{Goodisman}},
  \emph{\bibinfo{title}{Electrochemistry: Theoretical Foundations}}
  (\bibinfo{publisher}{John Wiley \& Sons}, \bibinfo{year}{1987}).

\bibitem{Tang:1992}
\bibinfo{author}{\bibfnamefont{Z.}~\bibnamefont{Tang}},
  \bibinfo{author}{\bibfnamefont{L.~E.} \bibnamefont{Scriven}},
  \bibnamefont{and} \bibinfo{author}{\bibfnamefont{H.~T.} \bibnamefont{Davis}},
  \bibinfo{journal}{J. Chem. Phys.} \textbf{\bibinfo{volume}{97}},
  \bibinfo{pages}{494} (\bibinfo{year}{1992}).

\bibitem{BoettingerReview:2002}
\bibinfo{author}{\bibfnamefont{W.~J.} \bibnamefont{Boettinger}},
  \bibinfo{author}{\bibfnamefont{J.~A.} \bibnamefont{Warren}},
  \bibinfo{author}{\bibfnamefont{C.}~\bibnamefont{Beckermann}},
  \bibnamefont{and} \bibinfo{author}{\bibfnamefont{A.}~\bibnamefont{Karma}},
  \bibinfo{journal}{Annu. Rev. Mater. Res.} \textbf{\bibinfo{volume}{32}},
  \bibinfo{pages}{163} (\bibinfo{year}{2002}).

\bibitem{McFaddenReview:2002}
\bibinfo{author}{\bibfnamefont{G.~B.} \bibnamefont{McFadden}},
  \bibinfo{journal}{Contemporary Mathematics} \textbf{\bibinfo{volume}{306}},
  \bibinfo{pages}{107} (\bibinfo{year}{2002}).

\bibitem{Harrowell:1987}
\bibinfo{author}{\bibfnamefont{P.~R.} \bibnamefont{Harrowell}}
  \bibnamefont{and} \bibinfo{author}{\bibfnamefont{D.~W.}
  \bibnamefont{Oxtoby}}, \bibinfo{journal}{J. Chem. Phys.}
  \textbf{\bibinfo{volume}{86}}, \bibinfo{pages}{2932} (\bibinfo{year}{1987}).

\bibitem{ElPhF:kinetics}
\bibinfo{author}{\bibfnamefont{J.~E.} \bibnamefont{Guyer}},
  \bibinfo{author}{\bibfnamefont{W.~J.} \bibnamefont{Boettinger}},
  \bibinfo{author}{\bibfnamefont{J.~A.} \bibnamefont{Warren}},
  \bibnamefont{and} \bibinfo{author}{\bibfnamefont{G.~B.}
  \bibnamefont{{McF}adden}}, \emph{\bibinfo{title}{Phase field modeling of
  electrochemistry II: Kinetics}}, \bibinfo{note}{unpublished},
  \eprint{cond-mat/0308179v1}.

\bibitem{Valette:1973}
\bibinfo{author}{\bibfnamefont{G.}~\bibnamefont{Valette}} \bibnamefont{and}
  \bibinfo{author}{\bibfnamefont{A.}~\bibnamefont{Hamelin}},
  \bibinfo{journal}{Electroanal. Chem. Interf. Electrochem.}
  \textbf{\bibinfo{volume}{45}}, \bibinfo{pages}{301} (\bibinfo{year}{1973}).

\bibitem{Valette:1982}
\bibinfo{author}{\bibfnamefont{G.}~\bibnamefont{Valette}}, \bibinfo{journal}{J.
  Electroanal. Chem.} \textbf{\bibinfo{volume}{138}}, \bibinfo{pages}{37}
  (\bibinfo{year}{1982}).

\bibitem{Gibbs:06}
\bibinfo{author}{\bibfnamefont{J.~W.} \bibnamefont{Gibbs}},
  \emph{\bibinfo{title}{The Scientific Papers of J. Willard Gibbs}}
  (\bibinfo{publisher}{Longmans, Green, and Co., London},
  \bibinfo{year}{1906}).

\bibitem{GBMcF:GibbsThomson}
\bibinfo{author}{\bibfnamefont{G.~B.} \bibnamefont{{McF}adden}},
  \emph{\bibinfo{title}{Electrochemical Gibbs-Thomson effect}},
  \bibinfo{note}{unpublished research}.

\bibitem{Wang:1993}
\bibinfo{author}{\bibfnamefont{S.-L.} \bibnamefont{Wang}},
  \bibinfo{author}{\bibfnamefont{R.}~\bibnamefont{Sekerka}},
  \bibinfo{author}{\bibfnamefont{A.~A.} \bibnamefont{Wheeler}},
  \bibinfo{author}{\bibfnamefont{B.~T.} \bibnamefont{Murray}},
  \bibinfo{author}{\bibfnamefont{S.~R.} \bibnamefont{Coriell}},
  \bibinfo{author}{\bibfnamefont{R.~J.} \bibnamefont{Braun}}, \bibnamefont{and}
  \bibinfo{author}{\bibfnamefont{G.~B.} \bibnamefont{{McF}adden}},
  \bibinfo{journal}{Physica D} \textbf{\bibinfo{volume}{69}},
  \bibinfo{pages}{189} (\bibinfo{year}{1993}).

\bibitem{Hart:1962}
\bibinfo{author}{\bibfnamefont{E.~J.} \bibnamefont{Hart}} \bibnamefont{and}
  \bibinfo{author}{\bibfnamefont{J.~W.} \bibnamefont{Boag}},
  \bibinfo{journal}{J. Amer. Chem. Soc.} \textbf{\bibinfo{volume}{84}},
  \bibinfo{pages}{4090} (\bibinfo{year}{1962}).

\bibitem{Egan:1985}
\bibinfo{author}{\bibfnamefont{J.~J.} \bibnamefont{Egan}} \bibnamefont{and}
  \bibinfo{author}{\bibfnamefont{W.}~\bibnamefont{Freyland}},
  \bibinfo{journal}{Ber. Bunsenges. Phys. Chem.} \textbf{\bibinfo{volume}{89}},
  \bibinfo{pages}{381} (\bibinfo{year}{1985}).

\bibitem{Wheeler:1992}
\bibinfo{author}{\bibfnamefont{A.~A.} \bibnamefont{Wheeler}},
  \bibinfo{author}{\bibfnamefont{W.~J.} \bibnamefont{Boettinger}},
  \bibnamefont{and} \bibinfo{author}{\bibfnamefont{G.~B.}
  \bibnamefont{{McF}adden}}, \bibinfo{journal}{Phys. Rev. A}
  \textbf{\bibinfo{volume}{45}}(\bibinfo{number}{10}), \bibinfo{pages}{7427}
  (\bibinfo{year}{1992}).

\bibitem{NumRec:C}
\bibinfo{author}{\bibfnamefont{W.~H.} \bibnamefont{Press}},
  \bibinfo{author}{\bibfnamefont{S.~A.} \bibnamefont{Teukolsky}},
  \bibinfo{author}{\bibfnamefont{W.~T.} \bibnamefont{Vetterling}},
  \bibnamefont{and} \bibinfo{author}{\bibfnamefont{B.~P.}
  \bibnamefont{Flannery}}, \emph{\bibinfo{title}{Numerical Recipes in {C}: the
  Art of Scientific Computing}} (\bibinfo{publisher}{Cambridge University
  Press}, \bibinfo{year}{1999}), 2nd ed.

\bibitem{VoGH84}
\bibinfo{author}{\bibfnamefont{D.}~\bibnamefont{Gottlieb}},
  \bibinfo{author}{\bibfnamefont{M.~Y.} \bibnamefont{Hussaini}},
  \bibnamefont{and} \bibinfo{author}{\bibfnamefont{S.~A.}
  \bibnamefont{Orszag}}, in \emph{\bibinfo{booktitle}{Spectral Methods for
  Partial Differential Equations}}, edited by
  \bibinfo{editor}{\bibfnamefont{R.~G.} \bibnamefont{Voigt}},
  \bibinfo{editor}{\bibfnamefont{D.}~\bibnamefont{Gottlieb}}, \bibnamefont{and}
  \bibinfo{editor}{\bibfnamefont{M.~Y.} \bibnamefont{Hussaini}}
  (\bibinfo{publisher}{SIAM, Philadelphia}, \bibinfo{year}{1984}), pp.
  \bibinfo{pages}{1--54}.

\bibitem{Pate84}
\bibinfo{author}{\bibfnamefont{A.~T.} \bibnamefont{Patera}},
  \bibinfo{journal}{J. Comp. Phys.} \textbf{\bibinfo{volume}{54}},
  \bibinfo{pages}{468} (\bibinfo{year}{1984}).

\bibitem{SNSQ}
\bibinfo{author}{\bibfnamefont{K.~L.} \bibnamefont{Hiebert}},
  \emph{\bibinfo{title}{\textsc{SLATEC} Common Math Library}},
  \bibinfo{organization}{National Energy Software Center},
  \bibinfo{address}{Argonne Natinal Laboratory, Argonne, {IL}}
  (\bibinfo{year}{1980}), \bibinfo{note}{based on Powell \cite{Powe70}}.

\bibitem{Bernard:2001}
\bibinfo{author}{\bibfnamefont{M.-O.} \bibnamefont{Bernard}},
  \bibinfo{author}{\bibfnamefont{M.}~\bibnamefont{Plapp}}, \bibnamefont{and}
  \bibinfo{author}{\bibfnamefont{J.-F.} \bibnamefont{Gouyet}}, in
  \emph{\bibinfo{booktitle}{Complexity and Fractals in Nature}}, edited by
  \bibinfo{editor}{\bibfnamefont{M.~M.} \bibnamefont{Novak}}
  (\bibinfo{publisher}{World Scientific, Singapore}, \bibinfo{year}{2001}), pp.
  \bibinfo{pages}{235--246}.

\bibitem{Bernard:2003}
\bibinfo{author}{\bibfnamefont{M.-O.} \bibnamefont{Bernard}},
  \bibinfo{author}{\bibfnamefont{M.}~\bibnamefont{Plapp}}, \bibnamefont{and}
  \bibinfo{author}{\bibfnamefont{J.-F.} \bibnamefont{Gouyet}},
  \emph{\bibinfo{title}{A mean-field kinetic lattice gas model of
  electrochemical cells}} (\bibinfo{year}{2003}), \eprint{cond-mat/0303072v1}.

\bibitem{Powe70}
\bibinfo{author}{\bibfnamefont{M.~J.~D.} \bibnamefont{Powell}}, in
  \emph{\bibinfo{booktitle}{Numerical Methods for Nonlinear Algebraic
  Equations}}, edited by
  \bibinfo{editor}{\bibfnamefont{P.}~\bibnamefont{Rabinowitz}}
  (\bibinfo{publisher}{Gordon and Breach, New York, {NY}},
  \bibinfo{year}{1970}), pp. \bibinfo{pages}{87--161}.

\bibitem{Cahn:1979}
\bibinfo{author}{\bibfnamefont{J.~W.} \bibnamefont{Cahn}}, in
  \emph{\bibinfo{booktitle}{Interfacial Segregation}}, edited by
  \bibinfo{editor}{\bibfnamefont{W.~C.} \bibnamefont{Johnson}}
  \bibnamefont{and} \bibinfo{editor}{\bibfnamefont{J.~M.}
  \bibnamefont{Blakely}} (\bibinfo{publisher}{ASM, Metals Park, {OH}},
  \bibinfo{year}{1979}), pp. \bibinfo{pages}{3--23}.

\end{thebibliography}
\bibliographystyle{apsrev}

\appendix

\section{First Integral}
\label{sec:FirstIntegral}

Here we characterize a first integral of the steady-state one-dimensional
equilibrium equations, and use it to obtain an expression for 
the surface \PRE{free} energy \SurfaceEnergy. 

It is convenient to introduce the electrochemical free energy density
\begin{equation}
    \ElectrochemicalPerVol
    \left(
	\Phase,\Concentration{1}\ldots\Concentration{\Components},\Potential
    \right) 
    = \sum_{j=1}^\Components \Concentration{j} 
    \Electrochemical{j}
    \left(
	\Phase,\Concentration{1}\ldots\Concentration{\Components},\Potential
    \right).
    \label{eq:HelmholtzPerVol:Electrochemical}
\end{equation}   
Here \( \Electrochemical{j} = \Chemical{j}{} + \Valence{j} \Faraday \Potential =
\partial \ElectrochemicalPerVol / \partial \Concentration{j} \) is the
electrochemical potential of species \( j \), and the charge density is given by
\( \ChargeDensity = \partial \ElectrochemicalPerVol / \partial \Potential \). 
The steady-state one-dimensional equilibrium equations can then be written
compactly in terms of \ElectrochemicalPerVol, and assume the form
\begin{subequations}
    \begin{align}
	\Electrochemical{j} - 
	\frac{\PartialMolarVolume{j}}{\PartialMolarVolume{\Substitutional}}
	\Electrochemical{\Solvent{}} 
	&= \Lagrange{j\Solvent{}},
	\qquad j = 1\ldots\Components-1
	\label{eq:Governing:Alternate:Electrochemical} 
	\\
	\frac{\partial \ElectrochemicalPerVol}{\partial \Phase} 
	- \Gradient{\Phase} \Phase_{\Position\Position} 
	- \frac{\Dielectric'(\Phase)}{2} \Potential_\Position^{2} 
	&= 0, 
	\label{eq:Governing:Alternate:Phase} 
	\\
	\frac{\partial \ElectrochemicalPerVol}{\partial \Potential} 
	+ \left( 
	    \Dielectric(\Phase) \Potential_\Position 
	\right)_\Position 
	&= 0.
	\label{eq:Governing:Alternate:Poisson}
    \end{align}
    \label{eq:Governing:Alternate}
\end{subequations}

If we differentiate the electrochemical free energy density of the equilibrium
solution with respect to \( \Phase(\Position) \), \( \Concentration{j}(\Position)
\), and \( \Potential(\Position) \), we find
\begin{multline}
    \frac{d}{d\Position}
    \ElectrochemicalPerVol
    \left(
	\Phase(\Position),
	\Concentration{1}(\Position)\ldots\Concentration{\Components}(\Position),
	\Potential(\Position)
    \right) 
    \\
    \begin{aligned}
	&= \frac{\partial \ElectrochemicalPerVol}{\partial \Phase} 
	    \Phase'(\Position) 
	+ \sum_{j=1}^\Components 
	    \frac{\partial \ElectrochemicalPerVol}{\partial \Concentration{j}} 
	    \Concentration{j}'(\Position)
	+ \frac{\partial \ElectrochemicalPerVol}{\partial \Potential} 
	    \Potential'(\Position)
	\\
	&= \Phase'(\Position) 
	\left[ 
	    \Gradient{\Phase} \Phase_{\Position\Position} 
	    + \frac{\Dielectric'(\Phase)}{2} \Potential_\Position^{2} 
	\right] 
	\\
	& \qquad 
	+ \sum_{j=1}^{\Components-1} 
	\left[ 
	    \Electrochemical{j} 
	    - \frac{\PartialMolarVolume{j}}{\PartialMolarVolume{\Substitutional}} 
	    \Electrochemical{\Solvent{}} 
	\right] \Concentration{j}'(\Position)
	- \Potential'(\Position) 
	\left( 
	    \Dielectric(\Phase) \Potential_\Position 
	\right)_\Position,
    \end{aligned}
\end{multline}
where we have used the volume constraint Eq.~\eqref{eq:VolumeConstraint} and the
governing equations \eqref{eq:Governing:Alternate} to eliminate the partial
derivatives of \ElectrochemicalPerVol.  This expression can be simplified to give
\begin{multline} 
    \frac{d}{d\Position} 
    \left[ 
	\ElectrochemicalPerVol
	\left(
	    \Phase(\Position),
	    \Concentration{1}(\Position)\ldots\Concentration{\Components}(\Position),
	    \Potential(\Position)
	\right) 
    \vphantom{\sum_{j=1}^{\Components-1} \Lagrange{j\Solvent{}} \Concentration{j}}
    \right.
    \\
    \left.
	- \sum_{j=1}^{\Components-1} \Lagrange{j\Solvent{}} \Concentration{j}
	- \frac{\Gradient{\Phase}}{2} \Phase_\Position^2 
	+ \frac{\Dielectric(\Phase)}{2} \Potential_\Position^2 
    \right] = 0.
    \label{eq:FirstIntegral}
\end{multline}
Since, from Eqs.~\eqref{eq:VolumeConstraint} and
\eqref{eq:HelmholtzPerVol:Electrochemical} we have
\begin{multline}
    \ElectrochemicalPerVol
    \left(
	\Phase(\Position),
	\Concentration{1}(\Position)\ldots\Concentration{\Components},
	\Potential(\Position)
    \right)
    \\
    \begin{aligned}
	&= \sum_{j=1}^{\Components-1} 
	\left[ 
	    \Electrochemical{j} - 
	    \frac{\PartialMolarVolume{j}}{\PartialMolarVolume{\Substitutional}} 
	    \Electrochemical{\Solvent{}} 
	\right] 
	\Concentration{j} 
	+ \frac{\Electrochemical{\Solvent{}}}{\PartialMolarVolume{\Substitutional}} 
	\\
	&= \sum_{j=1}^{\Components-1} \Lagrange{j\Solvent{}} \Concentration{j} 
	+ \frac{\Electrochemical{\Solvent{}}}{\PartialMolarVolume{\Substitutional}},
    \end{aligned}
\end{multline}
we may write the first integral represented by Eq.~\eqref{eq:FirstIntegral} in
the form
\begin{equation}
    \frac{\Electrochemical{\Solvent{}}}{\PartialMolarVolume{\Substitutional}} 
    - \frac{\Gradient{\Phase}}{2} \Phase_\Position^2 
    + \frac{\Dielectric(\Phase)}{2} \Potential_\Position^2 
    = \text{constant} 
    = \frac{\Electrochemical{\Solvent{}}^\infty}{\PartialMolarVolume{\Substitutional}},
    \label{eq:FirstIntegral:Equilibrium}
\end{equation}
where we have evaluated the integration constant in the far field
where \( \Phase_\Position = \Potential_\Position = 0 \) and \(
\Electrochemical{\Solvent{}} = \Electrochemical{\Solvent{}}^\infty \). 
In view of Eq.~\eqref{eq:Governing:Alternate:Electrochemical}, we
therefore find that the electrochemical potentials of the
substitutional species vary through the interface, with
\begin{equation}
   \Electrochemical{j} 
   = \Electrochemical{j}^\infty 
   + \PartialMolarVolume{j}
   \left(
       \frac{\Gradient{\Phase}}{2} \Phase_\Position^2 
      - \frac{\Dielectric(\Phase)}{2} \Potential_\Position^2
   \right).
   \qquad j = 1\ldots\Components
   \tag{\ref{eq:Electrochemical:Interface}}
\end{equation}
The interstitial species, with \( \PartialMolarVolume{j} = 0 \), thus have
uniform electrochemical potentials.

An alternative form of the free energy functional of
Eq.~\eqref{eq:HelmholtzFromElectrostatics} takes the form \footnote{Here we have
used the identity \( \int \ChargeDensity \Potential \, d\Position = \int
\Dielectric(\Phase) \Potential_\Position^2 \, d\Position \), which follows from
the Poisson equation with appropriate boundary conditions.}
\begin{multline}
     \Helmholtz
     \left(
	 \Phase,\Concentration{1}\ldots\Concentration{\Components},\Potential
     \right) 
     \\
     = \int_\Volume 
     \left[ 
	 \ElectrochemicalPerVol
	 \left(
	     \Phase,\Concentration{1}\ldots\Concentration{\Components},\Potential
	 \right) 
	 + \frac{\Gradient{\Phase}}{2}\abs{\nabla \Phase}^{2} 
	 - \frac{\Dielectric(\Phase)}{2}\abs{\nabla \Potential}^{2} 
     \right] d\Volume.
     \label{eq:Helmholtz:Electrochemical}
\end{multline}

\section{Surface Free Energy}
\label{sec:SurfaceEnergy}

A conventional definition of the surface \PRE{free} energy \SurfaceEnergy\ of a planar
interface at equilibrium between two isothermal, multicomponent, fluid phases,
with no electrical effects or volume constraints, is to write
\cite{Cahn:1979}
\begin{equation}
  \Helmholtz = \sum_{j=1}^\Components \Chemical{j}{\infty} \Number_j 
  - \Pressure^\infty \Volume 
  + \SurfaceEnergy \Area,
  \label{eq:Helmholtz:TwoPhase}
\end{equation}
where  \( \Pressure^\infty \) is the far field value of the pressure \Pressure. 
The interface is located in the interior of the region \( \BoxLeft <
\Position < \BoxRight \) and the free energy is
\begin{equation}
    \Helmholtz 
    = \Area \int_\BoxLeft^\BoxRight \HelmholtzPerVol \, d\Position
    = \Area \int_\BoxLeft^\BoxRight 
    \left(
	\sum_{j=1}^\Components \Chemical{j}{} \Concentration{j} - \Pressure
    \right)
     \, d\Position.
     \label{eq:Helmholtz:Generic}
\end{equation}

In our model of the electrolyte-electrode equilibrium, we are neglecting the
pressure term in Eqs.~\eqref{eq:Helmholtz:TwoPhase} and
\eqref{eq:Helmholtz:Generic}, and including a volume constraint and the
effects of an electric field on charged components.  The appropriate definition
of \SurfaceEnergy\ is analogous to Eq.~\eqref{eq:Helmholtz:TwoPhase}, with
\begin{equation}
  \Helmholtz 
  = \sum_{j=1}^\Components \Electrochemical{j}^\infty \Number_j
  + \SurfaceEnergy \Area,
\end{equation}
and \( \Number_{j} \) is defined by Eq.~\eqref{eq:MassConservation}.  The
surface \PRE{free} energy arises from the variation in the substitutional electrochemical
potentials across the interface.  From Eq.~\eqref{eq:Helmholtz:Electrochemical}
\begin{equation} 
    \SurfaceEnergy 
    = \int_\BoxLeft^\BoxRight
    \left[ 
	\ElectrochemicalPerVol 
	- \sum_{j=1}^\Components \Electrochemical{j}^\infty \Concentration{j} 
	+ \frac{\Gradient{\Phase}}{2} \Phase_\Position^2 
	- \frac{\Dielectric(\Phase)}{2} \Potential_\Position^2 
    \right] \, d\Position
    \label{eq:SurfaceEnergy:General}
\end{equation}
On substitution of Eqs.~\eqref{eq:HelmholtzPerVol:Electrochemical} and
\eqref{eq:Electrochemical:Interface} into Eq.~\eqref{eq:SurfaceEnergy:General}
we obtain
\begin{equation*}
    \SurfaceEnergy 
    = \int_\BoxLeft^\BoxRight
    \left[ 
	\Gradient{\Phase} \Phase_\Position^2 
	- \Dielectric(\Phase) \Potential_\Position^2 
    \right] \, d\Position.
    \tag{\ref{eq:SurfaceEnergy}}
\end{equation*}

\section{Surface Charge and Capacitance}
\label{sec:SurfaceCharge}

Here we derive the expression \eqref{eq:Adsorption:Classical} for the variation
in surface \PRE{free} energy \SurfaceEnergy\ associated with changes in the Galvani
potential \Galvani{\Standard} under the assumption of ideal solution
thermodynamics (Eq.~\eqref{eq:HelmholtzPerVolume}).  The derivation can be
generalized to non-ideal solution behavior if activity coefficients are
introduced.  The variation is computed for fixed values of the far-field mole
fractions \( \Fraction{j}^{\Electrode} \) and \( \Fraction{j}^{\Electrolyte} \),
so that from Eqs.~\eqref{eq:Electrochemical:Ideal} and
\eqref{eq:DeltaStandard:Standard} we see that the variation \( \Variation
\Galvani{\Standard} \) then induces corresponding variations \( \Variation
\Delta\Chemical{j}{\Standard} = -\Valence{j} \Faraday \Variation
\Galvani{\Standard} \) and the related expression \( \Variation
\Electrochemical{j}^\infty = -\Valence{j} \Faraday \Variation
\Galvani{\Standard} \).  The variation in \Galvani{\Standard} also leads to
variations \( \Variation \Potential(\Position) \), \( \Variation
\Concentration{j}(\Position) \), and \( \Variation \Phase(\Position) \) in the
equilibrium profiles of the field variables as well.  We compute the resulting
variation of the surface \PRE{free} energy, \( \Variation \SurfaceEnergy \).

From Eq.~\eqref{eq:SurfaceEnergy:General} we have
\begin{align}
    \Variation \SurfaceEnergy 
    &= \int_\BoxLeft^\BoxRight
    \left[ 
	\Variation \ElectrochemicalPerVol 
	- \sum_{j=1}^\Components \Electrochemical{j}^\infty 
	    \Variation \Concentration{j} 
	- \sum_{j=1}^\Components \Concentration{j} 
	    \Variation \Electrochemical{j}^\infty
	+ \Gradient{\Phase} \Phase_\Position \Variation \Phase_\Position 
    \right.
    \nonumber \\
    & \qquad \left.
    \vphantom{
	- \sum_{j=1}^\Components \Electrochemical{j}^\infty 
	    \Variation \Concentration{j} 
    }
	- \Dielectric(\Phase) \Potential_\Position 
	    \Variation \Potential_\Position 
	- \frac{\Dielectric'(\Phase)}{2} \Potential_\Position^2 
	    \Variation \Phase 
    \right] \, d\Position.
    \label{eq:Variation:SurfaceEnergy}
\end{align}
In computing the variation \( \Variation \ElectrochemicalPerVol \), we must
consider not only the explicit variations arising from \( \Variation
\Potential(\Position) \), \( \Variation \Concentration{j}(\Position) \), and \(
\Variation \Phase(\Position) \), but also take into account the implicit
variation associated with the dependence of \ElectrochemicalPerVol\ on
\( \Delta\Chemical{j}{\Standard} \).  We then find
\begin{align}
    \Variation \ElectrochemicalPerVol 
    &= \frac{\partial \ElectrochemicalPerVol}{\partial \Phase} 
	\Variation \Phase(\Position) 
    + \sum_{j=1}^\Components 
	\frac{\partial \ElectrochemicalPerVol}{\partial \Concentration{j}} 
	\Variation \Concentration{j}  
    \nonumber \\
    & \qquad + \frac{\partial \ElectrochemicalPerVol}{\partial \Potential} 
	\Variation \Potential(\Position) 
    + \sum_{j=1}^\Components 
	\frac{\partial \ElectrochemicalPerVol}
	    {\partial \Delta\Chemical{j}{\Standard}}
	\Variation \Delta\Chemical{j}{\Standard}
    \\
    &= \frac{\partial \ElectrochemicalPerVol}{\partial \Phase} 
	\Variation \Phase(\Position) 
    + \sum_{j=1}^\Components \Electrochemical{j} \Variation \Concentration{j}  
    \nonumber \\
    & \qquad + \frac{\partial \ElectrochemicalPerVol}{\partial \Potential} 
	\Variation \Potential(\Position) 
    - \Interpolate(\Phase) \ChargeDensity(\Position) 
	\Variation \Galvani{\Standard},
\end{align}
where we have used \( \partial \ElectrochemicalPerVol / \partial
\Delta\Chemical{j}{\Standard} = \Concentration{j} \Interpolate(\Phase) \)
and \( \Variation \Delta\Chemical{j}{\Standard} = -\Valence{j} \Faraday
\Variation \Galvani{\Standard} \).  Inserting this expression into
Eq.~\eqref{eq:Variation:SurfaceEnergy} and integrating by parts, we find
\begin{align} 
    \Variation \SurfaceEnergy 
    &= \int_\BoxLeft^\BoxRight
    \left\{ 
	\left[
	    \frac{\partial \ElectrochemicalPerVol}{\partial \Phase}
	    - \Gradient{\Phase} \Phase_{\Position\Position} 
	    - \frac{\Dielectric'(\Phase)}{2} \Potential_\Position^2 
	\right] \Variation \Phase 
    \right.
    \nonumber \\
    & \left. \qquad
	+ \sum_{j=1}^\Components 
	\left(
	    \Electrochemical{j} - \Electrochemical{j}^\infty 
	\right) \Variation \Concentration{j} 
	- \sum_{j=1}^\Components \Concentration{j} 
	    \Variation \Electrochemical{j}^\infty 
    \right.
    \nonumber \\
    & \left. \qquad {}  
	+ \left[ 
	    \frac{\partial \ElectrochemicalPerVol}{\partial \Potential} 
	    + \left( 
		\Dielectric(\Phase) \Potential_\Position 
	    \right)_\Position 
	\right] \Variation \Potential
	- \Interpolate(\Phase) \ChargeDensity(\Position) 
	    \Variation \Galvani{\Standard}
    \right\} \, d\Position.
\end{align}
Using the equilibrium equations \eqref{eq:Governing:Alternate},
Eq.~\eqref{eq:Electrochemical:Interface}, \( \Variation
\Electrochemical{j}^\infty = \Valence{j} \Faraday \Variation \Galvani{\Standard}
\), and \( \sum_{j=1}^\Components \PartialMolarVolume{j} \Variation
\Concentration{j} = 0 \) (which follows from Eq.~\eqref{eq:VolumeConstraint}),
to simplify the results, we obtain
\begin{equation}
    \Variation \SurfaceEnergy
    = \Variation \Galvani{\Standard} \int_\BoxLeft^\BoxRight
    \left[
	1 - \Interpolate(\Phase)
    \right] \ChargeDensity(\Position) \, d\Position,
    \label{eq:Adsorption:Variation}
\end{equation}
If we define the surface charge of the electrode as
\begin{align}
    \SurfaceCharge^\Electrode
    &\equiv \int_\BoxLeft^\BoxRight \Interpolate(\Phase) \ChargeDensity \, d\Position
    \tag{\ref{eq:SurfaceCharge:Electrode}}
\intertext{and the surface charge of the electrolyte as}
    \SurfaceCharge^\Electrolyte
    &\equiv \int_\BoxLeft^\BoxRight
    \left[
	1 - \Interpolate(\Phase)
    \right] \ChargeDensity \, d\Position,
    \label{eq:SurfaceCharge:Electrolyte}
\end{align}%
Eq.~\eqref{eq:Adsorption:Variation} recovers the classical electrochemical
adsorption formula of Eq.~\eqref{eq:Adsorption:Classical}.  Note that because
the total charge is zero, \( \SurfaceCharge^\Electrode =
-\SurfaceCharge^\Electrolyte \).

\section{Gouy-Chapman-Stern}
\label{sec:Classical:GouyChapmanStern}

It is useful to perform a detailed comparison to the standard Gouy-Chapman model
of the double layer.  This model only treats variations in the electrolyte and
the electrode-electrolyte interface is considered to be sharp.  The inputs to
the model are the difference between the voltage of the electrolyte at the metal
\( \Potential_{\IHP} \) and the voltage far from the interface \(
\Potential_{\infty} \), the dielectric constant, and the cation concentration of
the electrolyte far from the interface.  The Stern modification to the
Gouy-Chapman model requires an additional parameter \( \Position_{\OHP} \), the
location of the plane of closest approach to the electrode of ions with a finite
radius.  The model assumes a Boltzmann distribution in the electrolyte and
requires that Poisson's equation be satisfied, giving the voltage as a function
of distance from the metal into the electrolyte,
\begin{multline}
    \frac{\tanh
    \left[
	\Valence{\Cation{}}\Faraday
	\left(
	    \Potential-\Potential_{\infty}
	\right)
	/4\Gas\Temperature
    \right]
    }{\tanh
    \left[
	\Valence{\Cation{}}\Faraday
	\left(
	    \Potential_{\OHP}-\Potential_{\infty}
	\right)
	/4\Gas\Temperature
    \right]
    }
    = \exp
    \left[
	-(\Position-\Position_{\OHP})/\Thickness_{\Potential}^\text{GC}
    \right].
    \\
    0 < \Position_{\OHP} < \Position < \infty
    \label{eq:GC:potential}
\end{multline}
\Potential\ is linear for \( 0 < \Position < \Position_{\OHP} \) and \(
\Potential_{\OHP} \) is the potential at \( \Position_{\OHP} \) obtained by requiring
continuity of \Potential\ and of \( \nabla\Potential \) at \( \Position_{\OHP} \).
The Debye length of the system is
\begin{equation}
    \Thickness_{\Potential}^\text{GC}
    = 
    \left(
	    \frac{\Dielectric\Gas\Temperature}
		    {2\Concentration{\Cation{}}^{\infty}
			    \Valence{\Cation{}}^{2}\Faraday^{2}}
    \right)^{1/2}.
    \label{eq:GC:DebyeLength}
\end{equation}
From Gauss' law
\begin{equation}
    \SurfaceCharge^{\Electrode} = -\SurfaceCharge^{\Electrolyte}
    = -\Dielectric
    \left(
	\frac{d\Potential}{d\Position}
    \right)_{\Position=\Position_{\OHP}}
    \label{eq:SurfaceCharge:Gauss}
\end{equation}
and Eq.~\eqref{eq:GC:potential}, the surface charge in the metal as a function of \(
\Potential_{\OHP} \) is
\begin{equation}
    \SurfaceCharge^\Electrode = 
    \left(
	    8\Dielectric\Concentration{\Cation{}}^{\infty}
	\Gas\Temperature
    \right)^{1/2} 
    \sinh\frac{\Valence{\Cation{}}\Faraday
    \left(
	\Potential_{\OHP}-\Potential_{\infty}
    \right)
    }{2\Gas\Temperature}    
    \label{eq:GC:charge}
\end{equation}
and from Eq.~\eqref{eq:DifferentialCapacitance}, the differential capacitance
as a function of \( \Potential_{\OHP} \) is
\begin{equation}
    \frac{1}{\DifferentialCapacitance} = 
    \left[
        \left(
        \frac{2\Valence{\Cation{}}^{2}\Faraday^{2}
            \Dielectric\Concentration{\Cation{}}^{\infty}}
            {\Gas\Temperature}
        \right)^{1/2} 
        \cosh\frac{\Valence{\Cation{}}\Faraday
        \left(
        \Potential_{\OHP}-\Potential_{\infty}
        \right)
        }{2\Gas\Temperature}
    \right]^{-1}
    + \frac{\Position_{\OHP}}{\Dielectric}
    \label{eq:GC:capacitance}
\end{equation}

\end{document}